\documentclass[aps,pra,twocolumn,shortbibliography]{revtex4-1}
\usepackage{graphicx,amsmath,amssymb,mathtools}
\usepackage{dcolumn,bm}
\usepackage{physics}
\usepackage{dsfont,xcolor}
\usepackage[capitalise]{cleveref}

\newcommand{\xh}{\hat{x}}

\begin{document}

\title{1D quasicrystals and topological markers}
\author{Joseph Sykes}
\author{Ryan Barnett}
\affiliation{Department of Mathematics, Imperial College London, London SW7 2AZ, United Kingdom}

\begin{abstract}
    Local topological markers are effective tools for determining the topological properties of both homogeneous and inhomogeneous systems. The Chern marker is an established topological marker that has previously been shown to effectively reveal the topological properties of 2D systems. In an earlier work, the present authors have developed a marker that can be applied to 1D time-dependent systems which can be used to explore their topological properties, like charge pumping under the presence of disorder. In this paper, we show how to alter the 1D marker so that it can be applied to quasiperiodic and aperiodic systems. We then verify its effectiveness against different quasicrystal Hamiltonians, some which have been addressed in previous studies using existing methods, and others  which possess topological structures that have been largely unexplored. We also demonstrate that the altered 1D marker can be productively applied to systems that are fully aperiodic.
\end{abstract}

\maketitle

\section{introduction}

Topological quantum systems have become an increasingly active area of research over the past decade, due in part to the advancement of experimental techniques in cold atom and photonic systems. Topological systems posses potentially very useful properties, one of which is topological pumping in 1D systems where the charge passing through the system is quantized to a topological index \cite{thouless1983quantization}. The topological index of a system is generally characterized by the Chern number which is well defined for both 2D systems and 1D time dependent systems \cite{TKNNpaper}. The pumping behavior present in 1D time dependent systems has been investigated experimentally in both photonic and cold atomic gas systems \cite{verbin2015topological,wimmer2017experimental,lohse2016thouless,schweizer2016spin,nakajima2016topological,nakajima2021competition}.

There has also been a renewed interest in aperiodic and quasiperiodic systems as it has been shown that these systems can exhibit topological properties \cite{kraus2012equivalence, zilberberg2021topology,jagannathan2021fibonacci}. However, the traditional method of calculating the Chern number, as discussed in the TKNN paper \cite{TKNNpaper}, is only well defined for translationally invariant systems and therefore cannot be used to topologically classify quasicrystal systems or aperiodic systems. 

This problem has previously been circumvented by considering periodic approximations of quasicrystals and using the flux insertion method (see supplementary material of \cite{kraus2012equivalence}). Due to this approximation, large system sizes are required. With this one can then determine the Chern number of the system for gaps that remain open in the limit $L \to \infty$. Another way the topological structure of quasicrystals have been investigated is by considering the Aubry-Andr\'{e} model (also known as the Harper model) with an onsite potential which is incommensurate with the lattice spacing of the system \cite{aubry1980analyticity}. The parameter $b$ of the Aubry-Andr\'{e} model determines if the potential is commensurate with the lattice spacing. It was shown that this model can be used to model the topological properties of the Fibonacci quasicrystal when $b$ is set to the inverse of the golden ratio \cite{kraus2012equivalence}. The Aubry-Andr\'{e} model should be viewed as a special case due its bulk Hamiltonian being equivalent to the bulk Hamiltonian of the Harper-Hofstadter model which can be seen through dimensional reduction. As a result of this a Diophantine equation can be used to determine the Chern number of the Aubry-Andr\'{e} model, even if $b$ takes on an irrational value \cite{zilberberg2021topology}. The Diophantine equation, however, is only applicable to a specific set of topological models where the value of $b$ is known.

Both these techniques require one to either approximate the system or consider specific types of systems. This then highlights the need for a more general way of determining the topological index of 1D time dependent aperiodic systems which does not depend on the structure of a given model or approximations of the model and can be applied to any aperiodic time dependent topological model.

Another recent advance in the area of topological systems is that of topological markers. These markers allow one to calculate a system's topological index but, unlike the Chern number, do not require the system to be translationally invariant. As such, topological markers are ideal for investigating the topological indices of quasicrystals and systems with aperiodic ordering. One of the most prevalent topological markers is the Chern marker which is defined for 2D time independent systems and has previously been used to investigate the topological structure of inhomogeneous systems \cite{bianco2011mapping,Prodan2010Entanglement,tran2015topological,marrazzo2017locality,mitchell2018amorphous,Irsigler2019Interacting,Gerbert2020Local,ulcakar2020kibble,hayward2020effect,varjas2020computation}. The Chern marker has also been used to investigate the topological properties of 2D quasicrystals with different aperiodic tiling structures \cite{ghadimi2021topological,johnstone2021bulk}. While the Chern marker is defined for 2D systems it was recently shown by the authors of this paper 
that a topological marker can be defined for 1D time-dependent systems and be used to determine their topological index \cite{sykes2021local}. This method allowed for the possibility of spatial inhomogeneities that break translational symmetry, but relied on the notion of a well-defined unit cell once the spatial inhomogeneities are removed. For aperiodic systems the unit cell is ill defined whether the inhomogeneities are suppressed or not. As such, a modified topological marker is needed to investigate the topological properties of 1D time dependent quasicrystal and aperiodic systems.

In this paper we show that the existing 1D marker can be adapted such that it can be applied to both periodic and aperiodic topological systems. We demonstrate that this new marker correctly predicts the topological index of a system by first applying it to the Aubry-Andr\'{e} model, where the topological index is known, and then applying it to the Modified Rice-Mele model. For the latter model the TKNN form of the Chern number and the Diophantine equation cannot be used. We check the results for Modified Rice-Mele model by analysing the current through the system, checking it is quantised to the predicted topological index, along with analysing the shift in the Wannier centers in the bulk of the system.

The paper is structured as follows. \Cref{1D QC marker sec} reviews the 1D marker presented in \cite{sykes2021local} and defines a new 1D marker called the 1D quasicrystal marker, $M_{1Q}(x,t)$, which is well defined for aperiodic systems. \Cref{AA model sec} introduces the Aubry-Andr\'{e} model and applies $M_{1Q}(x,t)$ for both periodic and quasicrystal cases. It is shown that for both cases the change in $M_{1Q}(x,t)$ over a full time period correctly predicts the topological index of the system. \Cref{MRM sec} starts by introducing the Modified Rice-Mele model which allows an aperiodic sequence to be applied. \cref{sec: silver mean seq} then introduces the aperiodic silver mean sequence and applies it to the Modified Rice-Mele model showing that this system has a topological nature. $M_{1Q}(x,t)$ is then used to determine the topological index and is confirmed by analysing the current flowing through the bulk of the system as well as by analysing the shift in the Wannier centers of the system. It is also shown that considering the average behavior of a large group of bulk Wannier centers can be used to determine the system's topological index. However, the accuracy of this method is less than that of $M_{1Q}(x,t)$. \cref{sec: TM and PD sec} introduces the Thue-Morse and the period-doubling sequences which are aperiodic sequences but are not quasiperiodic like the Fibonacci sequence or the silver mean sequence. It is shown that topological gaps exist for the Modified Rice-Mele model with both the Thue-Morse sequence and the period-doubling sequence applied and in both cases $M_{1Q}(x,t)$ is used to predict the topological index of the system. \Cref{polarization sec} then shows that the evolution of $M_{1Q}(x,t)$ quantitatively matches the evolution of the polarization of the system without the discontinuous jump and that the evolution of the two are equivalent in the thermodynamic limit. Lastly, \cref{discus and conclus sec} reviews the paper and discusses the benefits of using the quasicrystal marker to determine the topological index. The differences between the time dependent Bott index used in \cite{yoshii2021charge} and $M_{1Q}(x,t)$ are also discussed along with the requirements a system needs to ensure that $M_{1Q}(x,t)$ correctly predicts the topological index.

\section{quasicrystal 1D marker} \label{1D QC marker sec}
In this section we will briefly review the 1D marker presented in \cite{sykes2021local} analysing each of the operators that make up the 1D marker. Where needed we will alter the definition of these operators such that they are well defined for both periodic and aperiodic systems. After this we will construct a new 1D marker which we label the 1D quasicrystal marker from these operators creating a topological marker that works for both periodic and aperiodic crystal structures.

It was shown in \cite{sykes2021local} that one can determine the topological index of a system with open boundary conditions by calculating the change in the 1D marker over a full time period within the bulk of the system. The 1D marker was defined as
\begin{equation} \label{1d marker}
    \mathcal{M}_1(x,t) = \frac{1}{V_c} {\rm tr}_{x} (\hat{U}^{\dagger} \, \hat{P} \, \hat{x} \,\hat{P} \, \hat{U})
\end{equation}
where $V_c$ is the volume of the unit cell, $\hat{P}$ projects into the occupied states of the Hamiltonian, $\hat{x}$ is the unit cell position operator, and $\hat{U}$ is the adiabatic evolution operator. To understand \cref{1d marker} we analyse the operators along with what is meant by ${\rm tr}_x(\hat O)$ (where $\hat{O}$ represents any operator) and adapt them to be well defined for aperiodic systems. 

The projector $\hat{P}$ is traditionally defined using the Bloch states of the Hamiltonian for translationally invariant systems. It is clear that this definition cannot be used for aperiodic systems and therefore we use an alternate form for the projector given by
\begin{equation}\label{projector}
    \hat{P} = \sum_{E \le E_F} \ket{\psi_E(t)}\bra{\psi_E(t)}
\end{equation}
where $\ket{\psi_E(t)}$ are the instantaneous eigenvectors of the Hamiltonian at time $t$ and $E_F$ is the Fermi-energy. This definition is well defined for both periodic and aperiodic systems. 

The adiabatic evolution operator $\hat{U}$ can be determined using the projectors in the following way. First one defines the effective adiabatic Hamiltonian, $\hat{h}$, given by
\cite{kato1950adiabatic, messiah1981quantum,kitaev2006anyons}
\begin{equation} \label{eff adiab ham}
    \hat{h} = i\;[\Dot{\hat{P}}\, ,\hat{P}]
\end{equation}
where $\dot{\hat{P}} = \frac{d}{dt} \hat{P}$ and the square brackets denote the commutator. After this, one solves the time evolution equation for this effective Hamiltonian, $i \frac{d}{dt} \hat{U} = \hat{h}\hat{U}$, with the initial condition $\hat{U} (t=0) = \mathds{1}$ to determine its evolution operator. This evolution operator evolves the system adiabatically and we therefore label it the adiabatic evolution operator. Using this operator one can then adiabatically evolve the projector in time using $\hat{P}(t) = \hat{U} \hat{\bar{P}} \hat{U}^\dagger$ where the over head bar indicates that the operator is evaluated at $t=0$. The adiabatic evolution operator, $\hat{U}$, is well defined for aperiodic systems when \cref{projector} defines the projector.

The position operator, $\hat{x}$, is defined as
\begin{equation}\label{x op}
    \hat{x} = \sum_x x\ket{x}\bra{x} \otimes \mathds{1}.
\end{equation}
In \cite{sykes2021local} $x$ labels the unit cell of the system and $\mathds{1}$ incorporates possible additional internal degrees of freedom like sites within a unit cell and spin. For systems with aperiodic structure the unit cell is ill defined and therefore we choose $x$ to label the position within the system and $\mathds{1}$ to label internal degrees of freedom of the system. In this paper we will focus on lattice systems with $x = an$,  where $a$ is the lattice spacing and $n$ is an integer which labels the lattice site position. We will also only consider spinless systems and therefore the $\mathds{1}$ can be dropped from \cref{x op}.

Lastly, the local trace defined in \cite{sykes2021local} was a trace that considered all the sites within a specific unit cell of the system. Again the unit cell is ill defined for aperiodic systems and therefore we will consider a trace over some region $\mathcal{R}$ of the system given by
\begin{equation}
    {\rm tr}_{\mathcal{R}}(\hat{O}) = \sum_{n \in \mathcal{R}} \bra{n}\hat{O}\ket{n}
\end{equation}
where $n$ labels the lattice sites of the system. In this work we will typically restrict the region $\mathcal{R}$ to be in the bulk of the system to ensure that we are calculating the behavior of the bulk. We also note that the prefactor $1 / V_c$ in \cref{1d marker} normalizes the trace by its size and as such we replace this with the normalization factor $1/L_{\mathcal{R}}$ where $L_{\mathcal{R}} = N_{\mathcal{R}}a$. Here $N_{\mathcal{R}}$ is the number of sites within the region $\mathcal{R}$ and $a$ is the size of the lattice spacing. Using the adjusted definitions of the operators and the trace, we now define the local 1D marker for quasicrystals as
\begin{equation}
    M_{1Q}(x,t) = \frac{1}{L_{\mathcal{R}}}{\rm tr}_{\mathcal{R}} \big(\hat{U}^\dagger\hat{P}\hat{x}\hat{P} \hat{U} \big).
\end{equation}
Throughout this paper we set the lattice spacing $a$ to 1 meaning that $L_{\mathcal{R}}=N_{\mathcal{R}}$. 

Now that we have defined the 1D marker for quasicrystals we will apply it to the well known Aubry-Andr\'{e} model for both the periodic and quasiperiodic cases.

\section{Aubry-Andr\'{e} model} \label{AA model sec}
Within this section we apply the 1D quasicrystal marker, $M_{1Q}(x,t)$, to the Aubry-Andr\'{e} model and show that the change in this value over a full time period gives the correct topological index of both periodic and aperiodic forms of this model. We use the Diophantine equation to confirm the topological index of the system.

The Aubry-Andr\'{e} (AA) model is an ideal model to evaluate $M_{1Q}(x,t)$ due to the fact that one can determine the topological index of this system when it possesses either a periodic or quasicrystaline nature. This is done using a Diophantine equation which is applicable due the AA model's close relationship with the Harper-Hofstadter model. It should be noted, however, that the Diophantine equation method only applies to select models like the Aubry-Andr\'{e} model and is not applicable in general. 

\begin{figure}
    \includegraphics[width=0.48\textwidth, trim = {0, 2cm, 0, 2cm}]{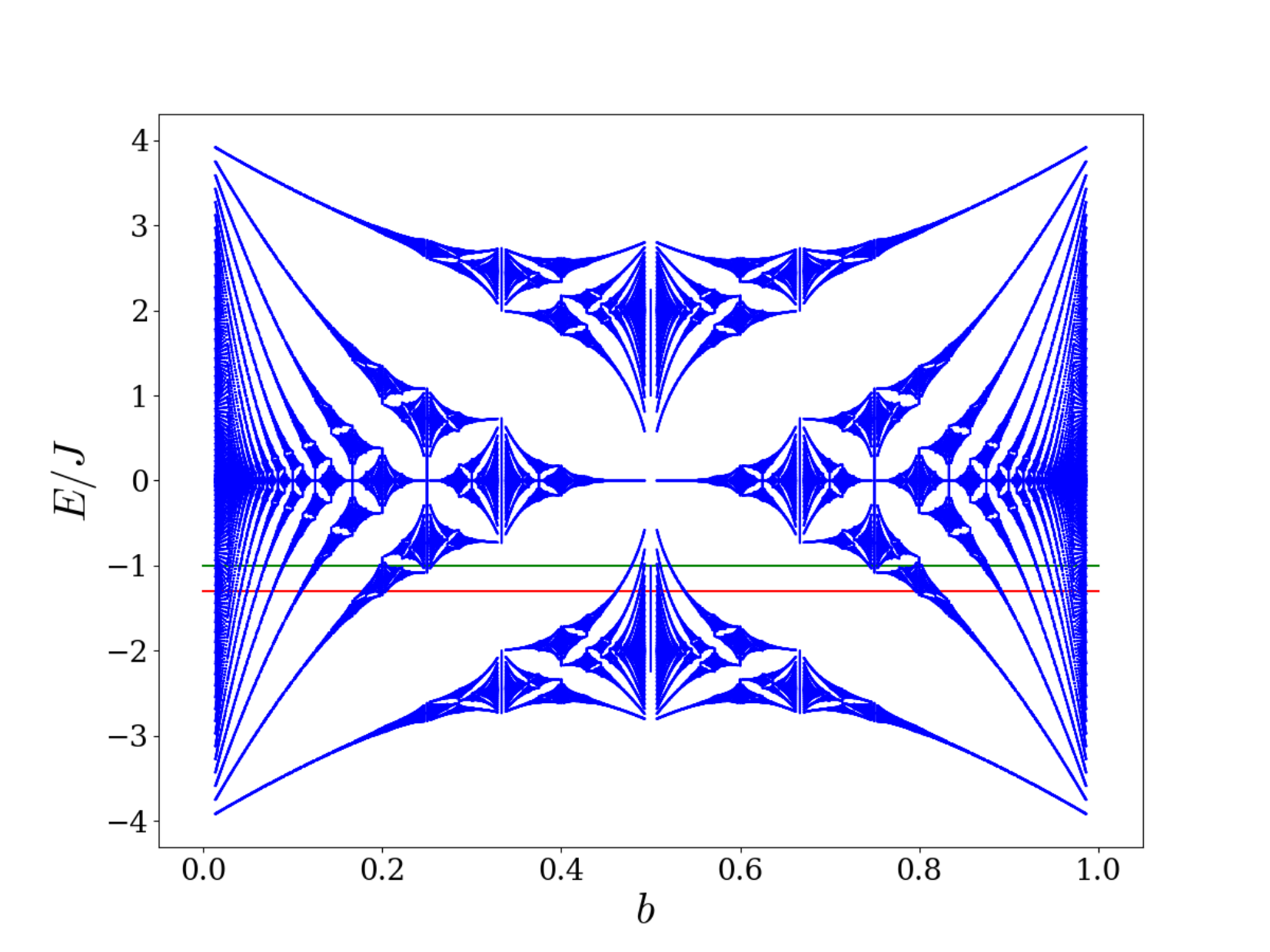}
    \caption{The Energy spectrum of the AA model for varying values of $b$. Here we set $\Delta/J = -2$. Lines at energies $E/J = -1$ and $E/J=-1.3$ are included to highlight the band gaps that exist at these energies for certain values of $b$.}
    \label{fig:Hofstadter_butterfly_for_AA_model}
\end{figure}

The Aubry-Andr\'{e} model is given by \cite{aubry1980analyticity}
\begin{equation} \label{AA model}
\begin{aligned}
    \hat{H}_{AA} = - J \sum_{n} & ( \ket{n}\bra{n+1}+ {\rm h.c.}) \\ &-\Delta \sum_{n}\text{cos}\big(2\pi b n - \phi(t)\big) \ket{n}\bra{n}
\end{aligned}
\end{equation}
where  $n$ is an integer labeling the lattice sites of the system, $J$ represents the strength of hoppings between sites and $\Delta$ represents the maximum magnitude of the on-site energy of the particles. The value $b$ determines if the system takes on a crystalline nature, achieved by setting $b$ to a rational fraction, or quasicrystalline nature, achieved by setting $b$ to an irrational fraction like the inverse of the golden ratio, $1 / \tau = 2/(1+\sqrt{5})$. The time-dependence is incorporated through $\phi (t)= 2\pi t/T$ where $T$ is the time period of the system. 

\Cref{fig:Hofstadter_butterfly_for_AA_model} shows the energy spectrum for the AA model with varying $b$. The parameters of the system were set to $\Delta / J = -2$ and periodic boundary conditions were taken. From this figure it can be seen that gaps exist for both $b=1/3$ and $b = 1/\tau \approx 0.618$. In this section we will consider the AA model with these values of $b$ and analyse the topological index of the system when the Fermi-energy is placed within these gaps.

It has been show that, due to its close relationship with the Harper-Hofstadter model, the topological index of the Aubry-Andr\'{e} model can be determined by the following Diophantine equation  \cite{bellissard1989spectral,jagannathan2021fibonacci,zilberberg2021topology}
\begin{equation}\label{altered diophantine eq}
    \rho_r = C_r b + t_r.
\end{equation}
Above, $\rho_r$ is the filling factor of the system, $b$ is the parameter of the AA model, $C_r$ is the Chern number of the system and $t_r$ is an integer chosen such that $C_r$ is also an integer. \Cref{altered diophantine eq} can also be referred to as the gap labeling theorem \cite{jagannathan2021fibonacci}.

We can now use $M_{1Q}(x,t)$ to predict the topological index of the AA model for the periodic and quasiperiodic cases and cross check it with \cref{altered diophantine eq}. We will first apply $M_{1Q}(x,t)$ to the periodic version of the AA model with $b=1/3$ showing that it gives the correct topological index of a periodic system. For this system the unit cell spans the length of three lattice sites.

\begin{figure}
    \includegraphics[width=0.48\textwidth, trim = {0, 2cm, 0, 2cm}]{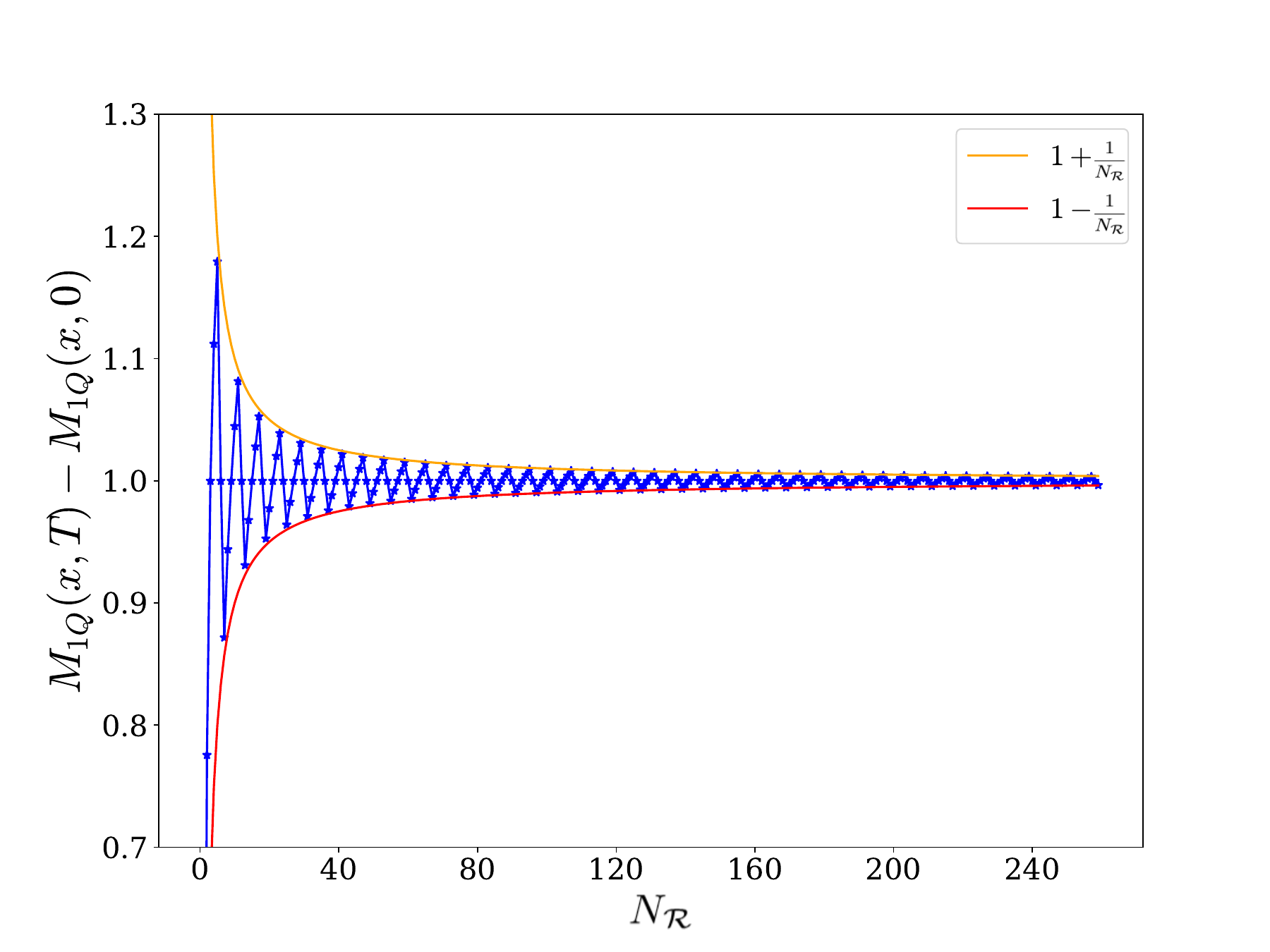}
    \caption{Change in $M_{1Q}(x,t)$ over a full time period for the Aubry-Andr\'{e} model with varying $N_{\mathcal{R}}$ and the parameter $b$ set to the rational value $b=1/3$. The other parameters of the system were set to  $\Delta/J = -2$ and $E_F / J =-1.3$. The total number of lattice sites within the system was set to $N=360$. The atomic spacing is normalized to unity. The red and orange lines highlight the envelope of the change in $M_{1Q}(x,t)$ as it tends to the Chern number of the system.}
    \label{fig:Change_in_M_1Q_for_varying_L_on_AA_model_with_b_is_1_over_3_FE_is_minus_1}
\end{figure}

\Cref{fig:Change_in_M_1Q_for_varying_L_on_AA_model_with_b_is_1_over_3_FE_is_minus_1} shows the change in $M_{1Q}(x,t)$ over a full time period for varying $N_{\mathcal{R}}$ with the parameters $\Delta/J = -2$ and a Fermi-energy of $E_F / J = -1.3$. It can be seen from \cref{fig:Hofstadter_butterfly_for_AA_model} that a gap exists at this Fermi-energy. The figure shows that when $N_{\mathcal{R}}$ equals an integer number of unit cells the change in $M_{1Q}(x,t)$ predicts the topological index of this gap to be one with a high degree of accuracy. Using \cref{altered diophantine eq}, one can show that for this Fermi-energy the value of $C_r$ is one. This then indicates that the change in $M_{1Q}(x,t)$ correctly predicts the Chern number of the system when $N_{\mathcal{R}}$ equals an integer number of unit cells. The figure also shows that the change in $M_{1Q}(x,t)$ tends to the value of one for increasing $N_{\mathcal{R}}$ and has an envelope of $1+1/N_{\mathcal{R}}$. Interestingly, it can be shown that this behavior persists when the denominator of $b$ takes on a large value. This then suggests that the behavior may also be present in quasicrystal systems seen as these systems can be approximated with rational values of $b$ with a large denominator.

To check if this behavior persists for irrational values of $b$ we next consider the AA model with $b = 1/ \tau$, where $\tau$ is the golden ratio. We set the Fermi-energy to $E_F /J = -1$ such that we lie within a large energy gap; all other parameters are set to those given in \cref{fig:Change_in_M_1Q_for_varying_L_on_AA_model_with_b_is_1_over_3_FE_is_minus_1}. For this case the AA model has the same topological structure as the Fibonacci quasicrystal model \cite{kraus2012equivalence}. \Cref{altered diophantine eq} has to be used to determine its topological index due to the TKNN form of the Chern number being ill defined for irrational values of $b$. Using \cref{altered diophantine eq}, one finds that the topological index of the system is $C_r = -1$.

\begin{figure}
    \includegraphics[width=0.48\textwidth, trim = {0, 2cm, 0, 2cm}]{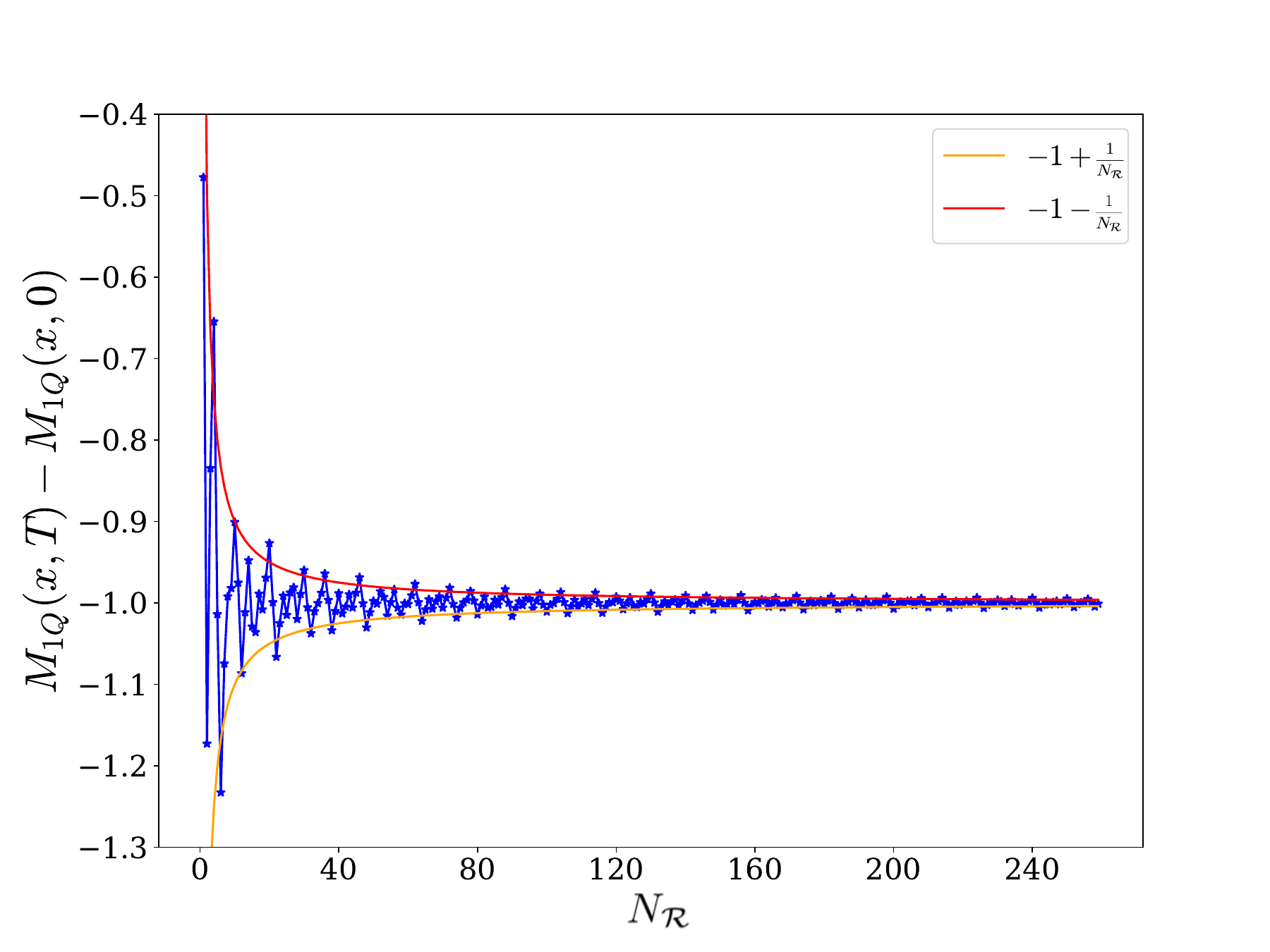}
    \caption{Change in $M_{1Q}(x,t)$ over a full time period for the Aubry-Andr\'{e} model with varying $N_{\mathcal{R}}$ and the parameter $b$ set to the irrational value $b=1/\tau$, where $\tau$ is the golden ration. The parameters were set to $\Delta/J = -2$ and $E_F / J = -1$. The total number of lattice sites within the system was set to $N=360$. The red and orange lines highlight the envelope of the change in $M_{1Q}(x,t)$ as it tends to the topological index of the system.}
    \label{fig:Change_in_M_1Q_for_varying_L_on_AA_model_with_b_is_1_over_golden_ratio_FE_is_minus_1}
\end{figure}

\cref{fig:Change_in_M_1Q_for_varying_L_on_AA_model_with_b_is_1_over_golden_ratio_FE_is_minus_1} shows that for the AA quasicrystal model the change in $M_{1Q}(x,t)$ over a full time period tends to the value of $C_r$ with increasing $N_{\mathcal{R}}$. One can also see that the envelope of the change in $M_{1Q}(x,t)$ approximately matches the one seen in the periodic case. When $N_{\mathcal{R}}=258$ the change in $M_{1Q}(x,t)$ equals $-1.0044$ giving the topological index of the system to two decimal places.

We also calculated the change in $M_{1Q}(x,t)$ for varying $N_{\mathcal{R}}$ for the AA quasicrystal with a Fermi-energy of $E_F / J = 2.18$. From \cref{fig:Hofstadter_butterfly_for_AA_model} one can again see that a gap exists at this Fermi-energy. Using \cref{altered diophantine eq}, the topological index can be shown to be $C_r = -2$. We found that in this case the change in $M_{1Q}(x,t)$ tended to the value of $-2$ and had an envelope that could be approximated by  $-2+2/N_{\mathcal{R}}$. This slightly differs from the behavior of the upper bound seen in \cref{fig:Change_in_M_1Q_for_varying_L_on_AA_model_with_b_is_1_over_golden_ratio_FE_is_minus_1}. 

Considering both these cases it is easy to see that the change in $M_{1Q}(x,t)$ tends to the Chern number of the system for increasing $N_{\mathcal{R}}$. We also observe by inspection that the envelope of the change in $M_{1Q}(x,t)$ can be approximated by
$C_r + \alpha/N_{\mathcal{R}}$ where $\alpha$ is a dimensionless constant of order unity. This then gives an approximate bound on the error of the change in $M_{1Q}(x,t)$ to be $\alpha/N_{\mathcal{R}}$ where the exact value of $\alpha$ is model specific.

We have thus shown that $M_{1Q}(x,t)$ can be used to predict the topological index of both periodic and quasicrystal systems. For the periodic case $N_{\mathcal{R}}$ should be set to an integer number of unit cells within the bulk to best predict the topological index. For the quasicrystal case $N_{\mathcal{R}}$ should be set to a large region in the bulk and will have an approximate bound on the error of $\alpha/N_{\mathcal{R}}$ with $\alpha$ being of order unity. Above, we considered a system where the Diophantine equation can be applied. Next, we will consider a system where the Diophantine equation cannot be applied and show that $M_{1Q}(x,t)$ correctly predicts the topological index of this system.

\section{Modified Rice-Mele Model} \label{MRM sec}
In what follows we introduce the Modified Rice-Mele model which allows the possibility to investigate systems with aperiodic structure and how this affects the topological structure of the system. There have been previous studies of inhomogeneity within Rice-Mele model using topological markers, but these studies have tended to asses inhomogeneity in the form of disorder and not quasicrystal structure \cite{hayward2020effect}. Of the studies where the effects of quasicrystal structures on the Rice-Mele model were investigated the concentration was on applying the Fibonacci sequence which has a well known topological structure \cite{yoshii2021charge}. Here we study the effects of the silver mean sequence, the Thue-Morse sequence, and the period-doubling sequence on the topological structure of the Rice-Mele model. The latter two sequences are aperiodic sequences but are not quasiperiodic in nature. 

At this point it is important to specify what a quasicrystal system is along with what an aperiodic system is and highlight the difference between the two. It is easiest to define an aperiodic system as a system that does not possess a periodic structure. A quasicrystal system is a special type of aperiodic system which possesses an almost periodic structure. Another way of distinguishing quasicrystal systems from other aperiodic system is by analyzing the Fourier transform of the lattice structure. When this is done a quasicrystal system will exhibit a countable set of peaks with a well-defined spacing \cite{kalish2018magnetoplasmonic}. It can be shown that a system generated using the silver mean sequence possesses a finite number of peaks in its Fourier transform and is therefore a quasicrystal. However, this is not the case for aperiodic systems that have a structure defined by the Thue-Morse sequence or the period-doubling sequence. As such, systems constructed using these sequences are aperiodic but not quasicrystals. For a more in-depth discussion see \cite{baake_grimm_2013}.

In \cref{sec: silver mean seq} we choose to apply the silver mean sequence to the system which is a quasiperiodic sequence and use $M_{1Q}(x,t)$ to predict the topological index of the system. To our knowledge there is no adaptation of the Diophantine equation to this model. As such, we check the topological index of the system by analysing the current in the bulk of the system as well as looking at the shift in the Wannier centers of the system. We also show that one can look at the average behavior of a group of Wannier centers within the bulk of the system over a time period and produce a single value that determines the topological index of the system. Comparing the results of this method to the results from the change in $M_{1Q}(x,t)$ over a time period we show that $M_{1Q}(x,t)$ provides a more accurate calculation of the topological index. In \cref{sec: TM and PD sec} we introduce the Thue-Morse sequence and the period-doubling sequence which are aperiodic sequences but are not quasiperiodic. We then apply these sequences to the Modified Rice-Mele model and use $M_{1Q}(x,t)$ to show that topological gaps appear for each sequence.

The Modified Rice-Mele (MRM) model is very similar to the original Rice-Mele (RM) model and was introduced in \cite{yoshii2021charge} where the Fibonacci sequence was applied. The Hamiltonian is defined as
\begin{equation}
\begin{aligned}
    &
    \begin{aligned}
    \hat{H}_{MRM} = \sum_n \bigg[ \Big[ \big(J + (-1)^{f_n} &\delta(t) \big) \ket{n}\bra{n+1} + h.c. \Big] \\
    &+ (-1)^{f_n} \gamma(t) \ket{n}\bra{n} \bigg],
    \end{aligned}
    \\
    \\
    &\quad \delta(t) = \delta_0\; {\rm cos}(2\pi t/T ) \quad\quad \gamma(t) = \gamma_0 \; {\rm sin}(2\pi t/T ).
\end{aligned}
\end{equation}
Here $n$ labels the lattice sites of the system, $J$ is a fixed hopping value and $\delta(t)$ and $\gamma(t)$ are time dependent modulating components for the hopping and on-site potential respectively. The modulating components both have the same period $T$. The key difference between the RM model and the MRM model is the value $f_n$. For the RM model $f_n = n$ which gives the system a periodic nature, however, the MRM model sets $f_n$ to equal some aperiodic sequence of $0$'s and $1$'s. For our analysis we choose to set $f_n$ to the silver mean sequence which we define in the next section.

\subsection{Silver mean sequence} \label{sec: silver mean seq}

One way to define an aperiodic sequence is through the substitution method where the initial value is given along with a number of substitution rules depending on how many different variables can be in the sequence. For the silver mean sequence the initial value is set to $A$ and the substitution rules are given by $g(A) = AAB$ and $g(B) = A$. Using this method the first four generations of the silver mean sequence are given by
\begin{equation}
    \begin{aligned}
    & S_0 = A \\
    & S_1 = AAB \\
    & S_2 = AABAABA \\
    & S_3 = AABAABAAABAABAAAB
    \end{aligned}
\end{equation}
where the subscript on $S$ is the generation number of the sequence. This sequence is a quasiperiodic sequence and the ratio of $A$'s to $B$'s tends to the value of the silver mean ratio, $\upsilon = 1+\sqrt{2}$, as the length of the sequence tends to infinity \cite{baake_grimm_2013}. To apply this sequence to our model we need to convert the sequence of $A$'s and $B$'s to a sequence of $0$'s and $1$'s. We do this by using the following rule
\begin{equation} \label{binary conversion}
\begin{aligned}
    f_n =  
    \begin{cases}
    1 \text{ if the } n\text{th letter of } S_i \text{ is } A\\
    0 \text{ if the } n\text{th letter of } S_i \text{ is }B.
    \end{cases}
\end{aligned}
\end{equation}

\begin{figure}
    \includegraphics[width=0.48\textwidth, trim = {0, 2cm, 0, 2cm}]{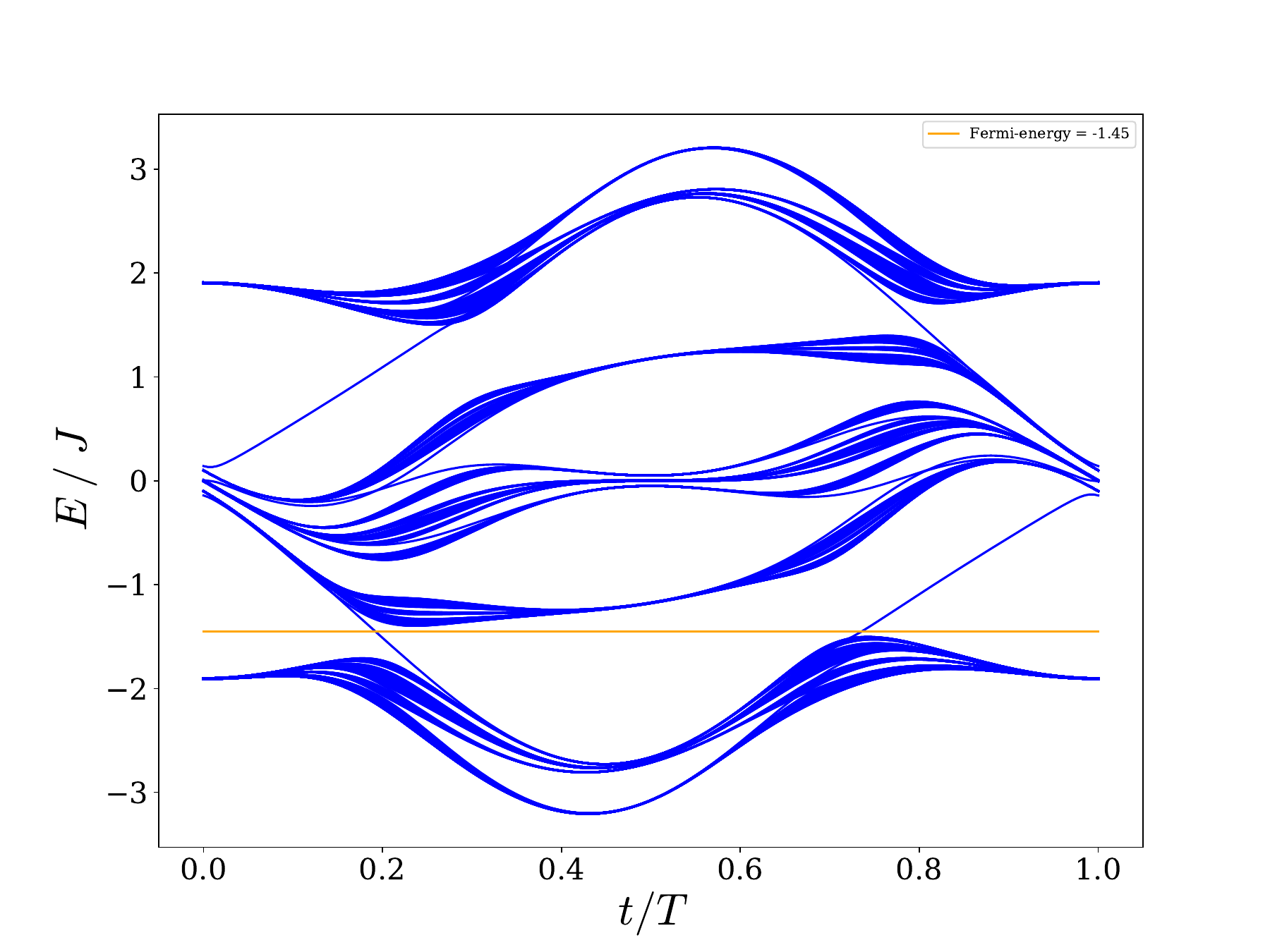}
    \caption{Energy spectrum of the Modified Rice-Mele model with open boundary conditions and the silver mean sequence applied. The parameters of the model were set to $\delta_0 /J = 0.9$ and $\gamma_0 /J = 0.8$. }
    \label{fig:Energy_spectrum_of_MRM_with_J_1_delta0_point9_gamma0_point8}
\end{figure}

\Cref{fig:Energy_spectrum_of_MRM_with_J_1_delta0_point9_gamma0_point8} shows the energy spectrum of the MRM model with open boundary conditions  and parameters $\delta_0 / J=0.9$ and $\gamma_0 / J=0.8$. From this figure it can be seen that two eigenvectors traverse the energy gap around $E/J=-1.45$ in the opposite direction. It can be shown that these eigenvectors are exponentially localised at the edges of the system suggesting that they may be topological in nature. If this is the case then, from the bulk boundary correspondence, the system will have a topological index of magnitude one when the Fermi-energy is set to $E_F / J = -1.45$ \cite{asboth2016short}.

\begin{figure}
    \includegraphics[width=0.48\textwidth, trim = {0, 2cm, 0, 2cm}]{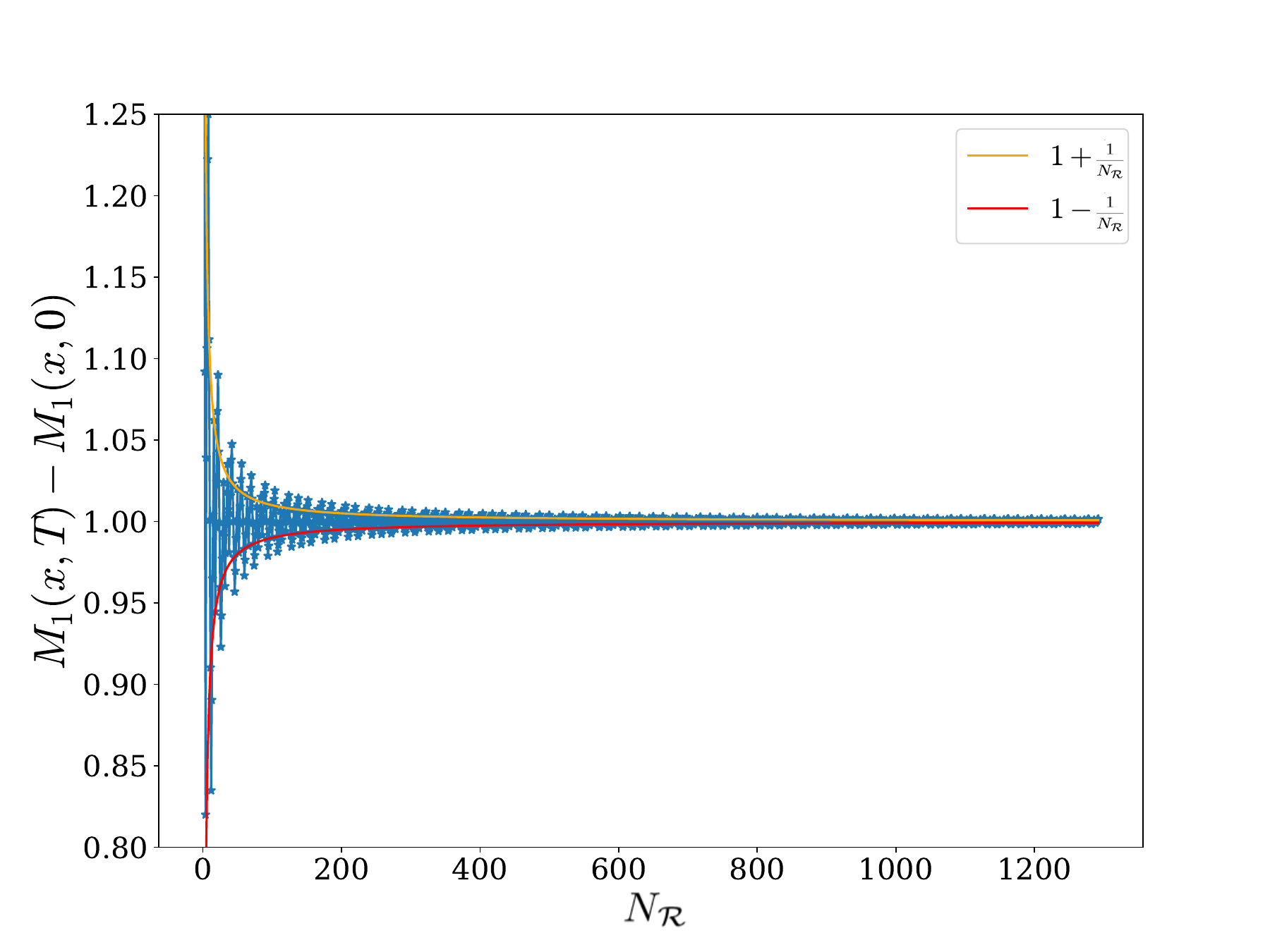}
    \caption{Change in $M_{1Q}(x,t)$ over a full time period for the Modified Rice-Mele model with the silver mean sequence applied and varying $N_{\mathcal{R}}$. The parameters were set to $\delta_0 / J = 0.9$, $\gamma_0 / J = 0.8$ and $T=1$. The Fermi-energy was set to $E_F / J = -1.45$. The total number of lattice sites within the system was set to $N=1393$. The red and orange lines highlight the envelope of the change in $M_{1Q}(x,t)$ as it tends to the topological index of the system.}
    \label{fig:Change_in_M_1Q_for_varying_L_on_MRM_model_with_delta_0point9_gamma_0point8_FE_is_minus_1point45}
\end{figure}

\Cref{fig:Change_in_M_1Q_for_varying_L_on_MRM_model_with_delta_0point9_gamma_0point8_FE_is_minus_1point45} shows the change in $M_{1Q}(x,t)$ over a full time period for the RM model with varying sizes of $N_{\mathcal{R}}$. The parameters are the same as in \cref{fig:Energy_spectrum_of_MRM_with_J_1_delta0_point9_gamma0_point8} and the Fermi-energy was set to $E_F / J = -1.45$. We see that the change in $M_{1Q}(x,t)$ tends to the value $1$ suggesting that this is the topological index of the system. This agrees with our previous prediction using bulk boundary correspondence that the magnitude of the topological index would be one. It can also be seen that the change in $M_{1Q}(x,t)$ possesses the same approximate $C_r\,+\,\alpha/N_{\mathcal{R}}$ envelope seen in previous models. This then suggests that approximate bound on the error of the change in $M_{1Q}(x,t)$ is given by $\alpha/N_{\mathcal{R}}$ for all periodic and aperiodic models with the value of $\alpha$ being model specific.

To confirm the system's topological index we use the fact that 1D topological systems pump an integer amount of charge through the bulk over a full time period equal to the system's topological index \cite{thouless1983quantization}. We will assess this bulk pumping by dividing the system into two sections and observing how the particle number changes in each half of the system over a full time period. We also confirm the topological index by analysing the shift in the Wannier centers given by the eigenvalues of $\hat{P}\hat{x}\hat{P}$. We show that the number of Wannier centers passing through a point in the system over a full time period is equal to the topological index of the system.

\begin{figure}
    \includegraphics[width=0.48\textwidth, trim = {0, 2cm, 0, 2cm}]{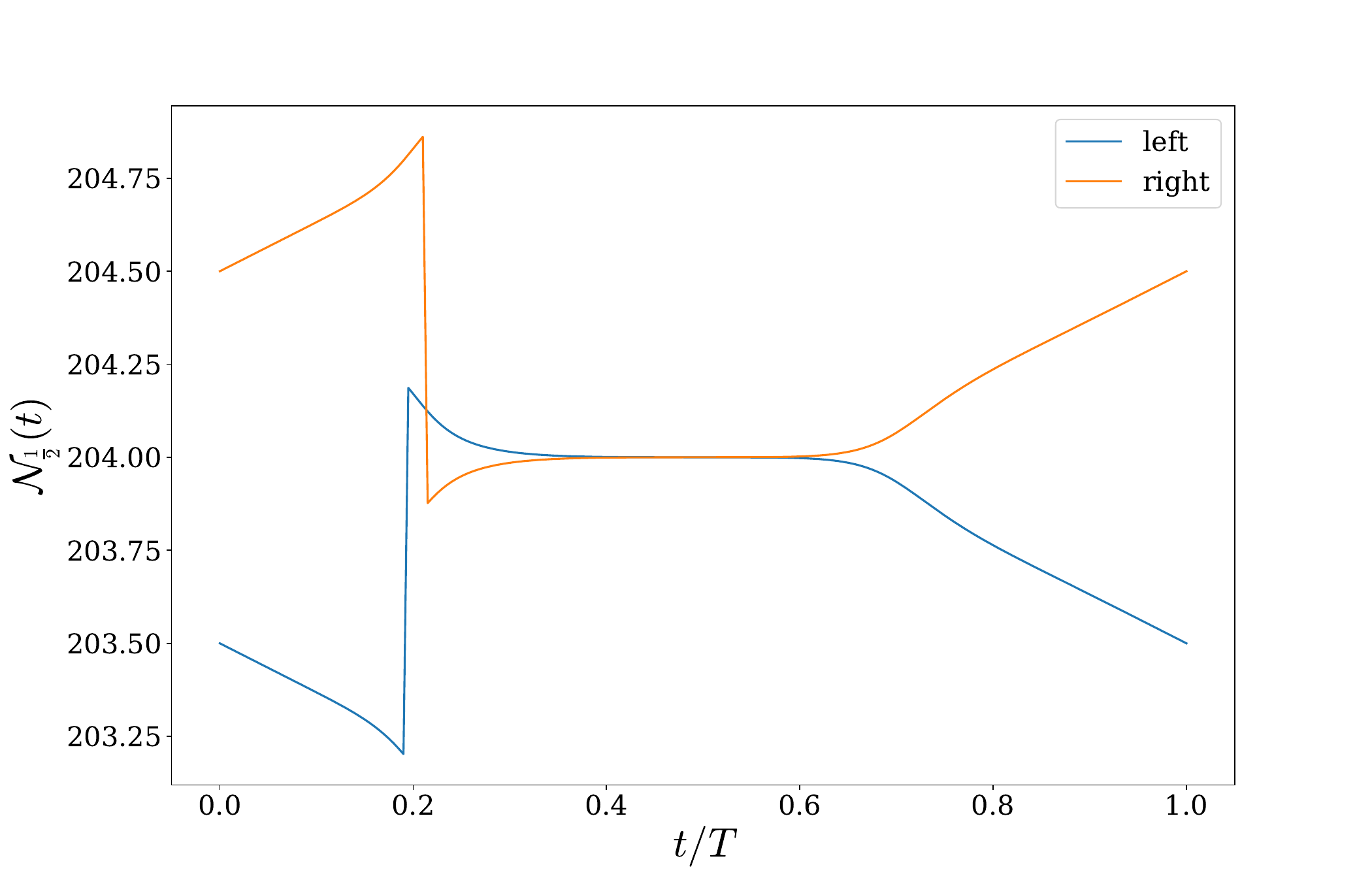}
    \caption{The change in the particle number in each half of the system over time for the Modified Rice-Mele model with the silver mean sequence applied. The parameters of the system are the same as those in \cref{fig:Change_in_M_1Q_for_varying_L_on_MRM_model_with_delta_0point9_gamma_0point8_FE_is_minus_1point45}. The light blue line represents the change in the particle number on the left side of the system and the orange line represents the change in the particle number on the right side of the system.}
    \label{fig:Change_in_density_on_eithersize_of_divide_for_MRM_with_delta_0point9_gamma_0point8_FE_minus_1point45}
\end{figure}

\Cref{fig:Change_in_density_on_eithersize_of_divide_for_MRM_with_delta_0point9_gamma_0point8_FE_minus_1point45} shows how the particle number in each half of the system changes over time. The particle number for the left hand side of the system was calculated in the following way
\begin{equation}
    \mathcal{N}_{LHS}(t) = \sum_{n \in LHS} \bra{n}\hat{P}\ket{n}.
\end{equation}
where $n$ labels the lattice sites of the system. The particle number for the right side of the system was calculated in a similar way. The discontinuous jumps in this figure occur where the Fermi-energy intersects with the edge state modes. These jumps are present because we disclude the edge state modes from the calculation of the particle number once their energy is above the Fermi-energy level which then leads to an integer change in the particle number. Because the discontinuous jumps represent edge state effects we can ignore them as we are only interested in bulk effects. Ignoring the discontinuities, it can be seen that the particle number for the left of the system is decreasing over time and over a full time period decreases by one; the particle number on the right of the system is the reverse, increasing by one over a full time period. This shows that an integer amount of charge flows across the divide in the system over a time period. It can also be shown that, as long as the partition lies within the bulk of the system, it does not matter where the system is divided, there will always be an integer amount of charge that passes through the partition over a full time period. We have therefore shown that one unit of charge flows through the bulk over a full time period, thus confirming that the topological index of the system is one and supporting the result we found using $M_{1Q}(x,t)$.

We now move on to show the shift in the Wannier centers of the system. It has previously been shown that one can define localised Wannier centers of both periodic and aperiodic 1D systems from the eigenvalues of $\hat{P}\hat{x}\hat{P}$ where $\hat{P}$ is given by \cref{projector} \cite{kivelson1982wannier,niu1991theory,nenciu1998existence}. In the periodic case one can consider the change in a single Wannier center deep within the bulk to determine the topological index of the system, however, this is not the case for aperiodic systems. In this case the aperiodic nature allows the Wannier centers to change by different amounts over a full time period rendering the method used in the periodic case unfruitful. However, we show here that the topological index of the system can be determined by analysing how many Wannier centers cross a horizontal line over a given period. One could ask whether the position of the line is important; we will show that it is not now. The periodic nature in time of the Hamiltonian, $\hat{H}(T) = \hat{H}(0)$, ensures that the final position of a Wannier center in the bulk is equal to the initial position of the next Wannier center. This means that it does not matter where one places the horizontal line as the number of Wannier centers that cross it over a full time period will remain the same as long as it is placed within the bulk. The horizontal line corresponds to a specific point in the system.

\begin{figure}
    \includegraphics[width=0.48\textwidth, trim = {0, 2cm, 0, 2cm}]{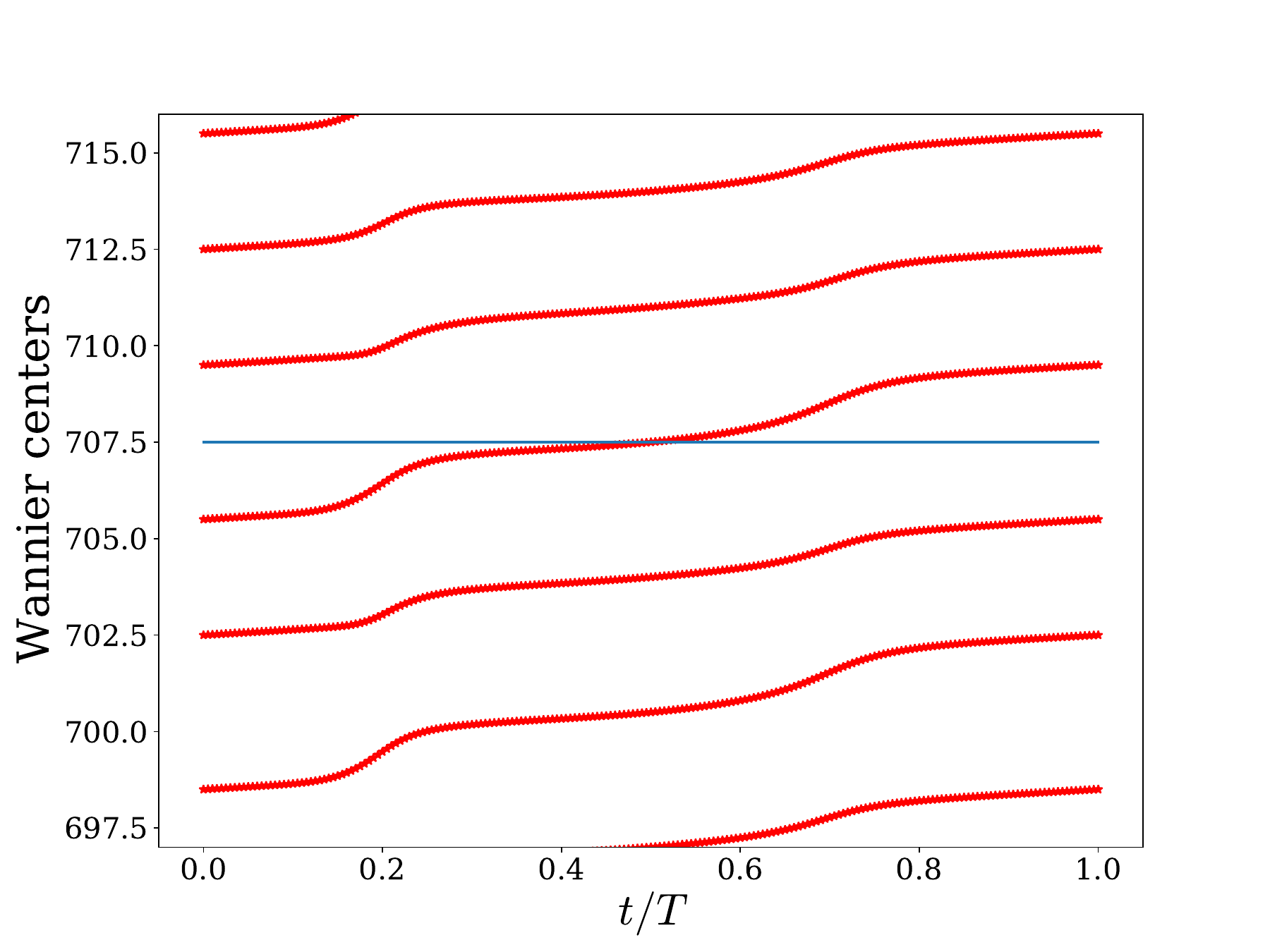}
    \caption{The time evolution of a group of Wannier centers in the bulk of the system over a full time period. The system considered was the MRM with the silver mean sequence applied. The parameters of the system are the same as in \cref{fig:Change_in_M_1Q_for_varying_L_on_MRM_model_with_delta_0point9_gamma_0point8_FE_is_minus_1point45}. The horizontal blue line represents a divide in the system located half way between the 707th and 708th lattice site.}
    \label{fig:Wannier_centres_in_bulk_for_MRM_model_with_delta_0point9_gamma_0point8_FE_is_minus_1point45}
\end{figure}

\Cref{fig:Wannier_centres_in_bulk_for_MRM_model_with_delta_0point9_gamma_0point8_FE_is_minus_1point45} shows the time evolution of a group of Wannier centers located in the bulk of the system. The horizontal light blue line represents a divide in the system placed between the 707th and 708th lattice sites. From this it can be seen that one Wannier center passes up through the Horizontal line indicating that the topological index is one. This agrees with both $M_{1Q}(x,t)$ and the current through the system. 

Summarizing this section so far, we have shown that $M_{1Q}(x,t)$ correctly predicts the topological index of the 1D time dependent MRM model. We confirmed the value of the topological index first by analysing the current through the system and showing it is quantised to the predicted topological value, and then by analysing the shift in the Wannier centers of the system. To the authors knowledge the Wannier center method presented above has not been used to determine the topological index of an aperiodic system before.

One could then ask why $M_{1Q}(x,t)$ should be used to determine the topological index rather than the Wannier center method or the current method show above. One advantage $M_{1Q}(x,t)$ has over the current method is that it does not contain discontinuous jumps in it evolution in time which can become cumbersome for higher topological indices. As well as this, $M_{1Q}(x,t)$ does not require visual analysis to determine the topological index of the system, unlike the Wannier center method presented above. However, one of the most important advantages of $M_{1Q}(x,t)$ is that it can be generalized to 3D systems using the results found in \cite{sykes2021local}. We are unaware of a generalization of the current method to 3D systems and Wannier centers are only necessarily localized for 1D systems.


\begin{figure}
    \includegraphics[width=0.48\textwidth, trim = {0, 2cm, 0, 2cm}]{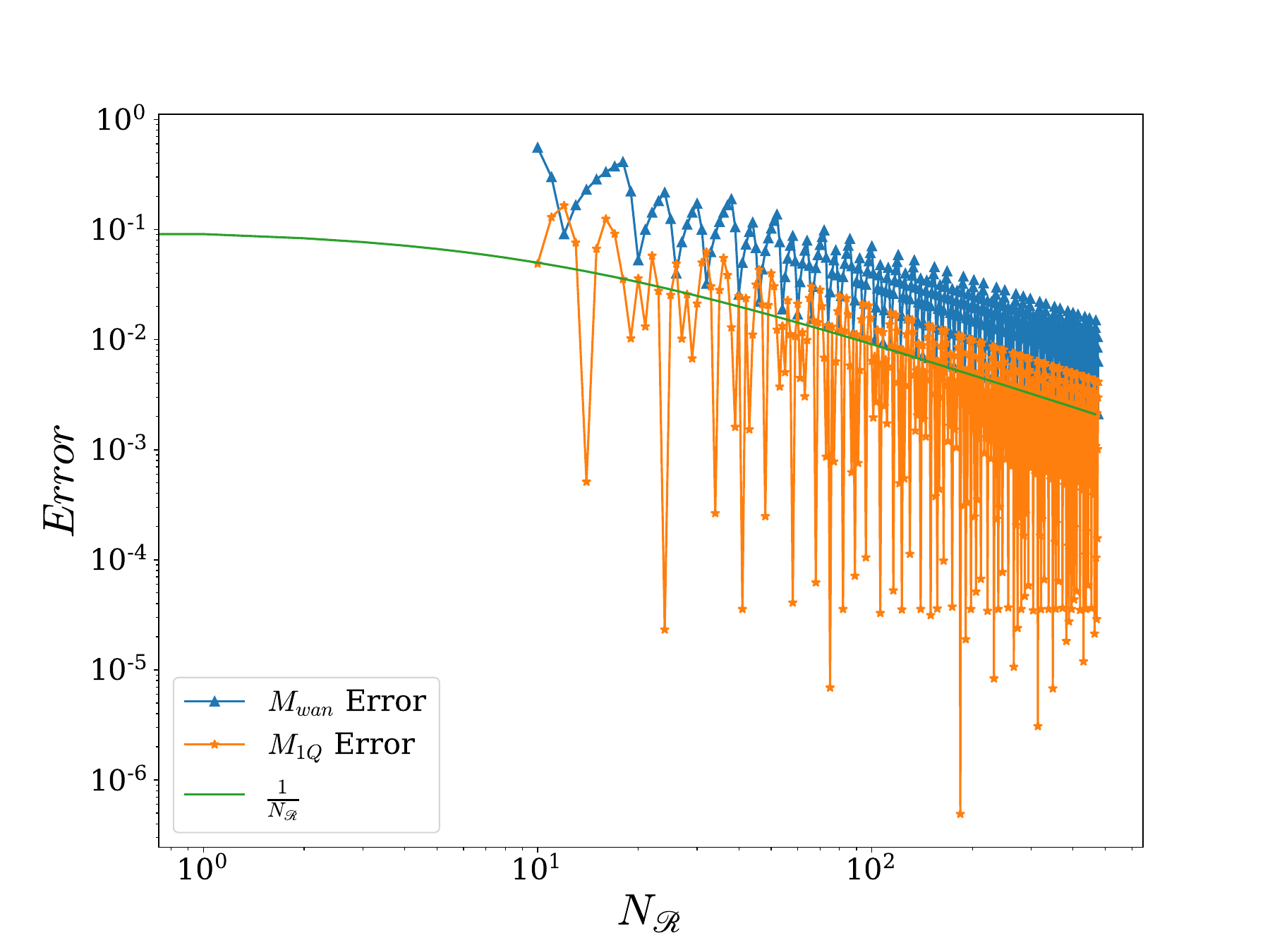}
    \caption{The error for both the change in $M_{1Q}(x,t)$ and $M_{wan}(t)$ compared to the true topological index for varying region size $N_{\mathcal{R}}$. The Wannier centers considered in $M_{wan}(t)$ were selected such that their initial positions at time $t=0$ lay within the region $\mathcal{R}$. The system considered was the MRM with the silver mean sequence applied. The parameters of the system are the same as in \cref{fig:Change_in_M_1Q_for_varying_L_on_MRM_model_with_delta_0point9_gamma_0point8_FE_is_minus_1point45}. The green line plots $1/N_{\mathcal{R}}$ and shows that the error for both $M_{1Q}(x,t)$ and $M_{wan}(t)$ is of order $\mathcal{O}(1/N_{\mathcal{R}})$.}
    \label{fig:Error_on_M_1Q_and_avg_Wannier_centre_for_MRM_with_delta_0point9_gamma_0point8_FE_minus_1point45}
\end{figure}

It is possible to consider the average of a group of Wannier centers located in a region $\mathcal{R}$ in the bulk to gain an overall picture of the systems bulk behavior. From this one can then calculate its change over time and normalize it such that it give a single value that can determine the system's topological index. However, as we will show, this method is less accurate than using $M_{1Q}(x,t)$. One can consider the average of the bulk Wannier centers normalized in the following way
\begin{equation}
    M_{wan}(t) = \frac{1}{W_{M+1}(0) - W_1(0)} \; \sum_{m = 1}^M W_m(t)
\end{equation}
where $W_m$ represents the Wannier centers and the summation considers Wannier centers positioned within the region $\mathcal{R}$ at time $t=0$. Calculating the change in $M_{wan}(t)$ over a full time period will then give you a single value that can be used to determine the topological index of the system.

\Cref{fig:Error_on_M_1Q_and_avg_Wannier_centre_for_MRM_with_delta_0point9_gamma_0point8_FE_minus_1point45} shows how the error for the change in $M_{1Q}(x,t)$ compares to the error for the change in $M_{wan}(t)$ for increasing region size, $N_{\mathcal{R}}$. It can be seen from this figure that the error on both these values is of order $\mathcal{O}(1/N_{\mathcal{R}})$,  highlighted by the green line. It can also be seen that the error of $M_{1Q}(x,t)$ is generally lower than that of $M_{wan}(t)$. Note that the axes of \cref{fig:Error_on_M_1Q_and_avg_Wannier_centre_for_MRM_with_delta_0point9_gamma_0point8_FE_minus_1point45} are logarithmic. As such, whilst one can produce a single value that determines the topological index from the Wannier centers it is less accurate than $M_{1Q}(x,t)$.

Another advantage of $M_{1Q}(x,t)$ is that one can use it to determine the topological index of different parts of the system and investigate how localized disorder affects the topological index in different regions of the system. We carried this out by separating the system into 10 sections and placed a local defect in one of these regions. Doing this we found that the topological index  in the region where the local disorder was placed was affected, but the topological index of the other regions remained unaffected by the local disorder.

\subsection{Thue-Morse and period-doubling sequence} \label{sec: TM and PD sec}

\begin{figure}
    \includegraphics[width=0.48\textwidth, trim = {0, 2cm, 0, 2cm}]{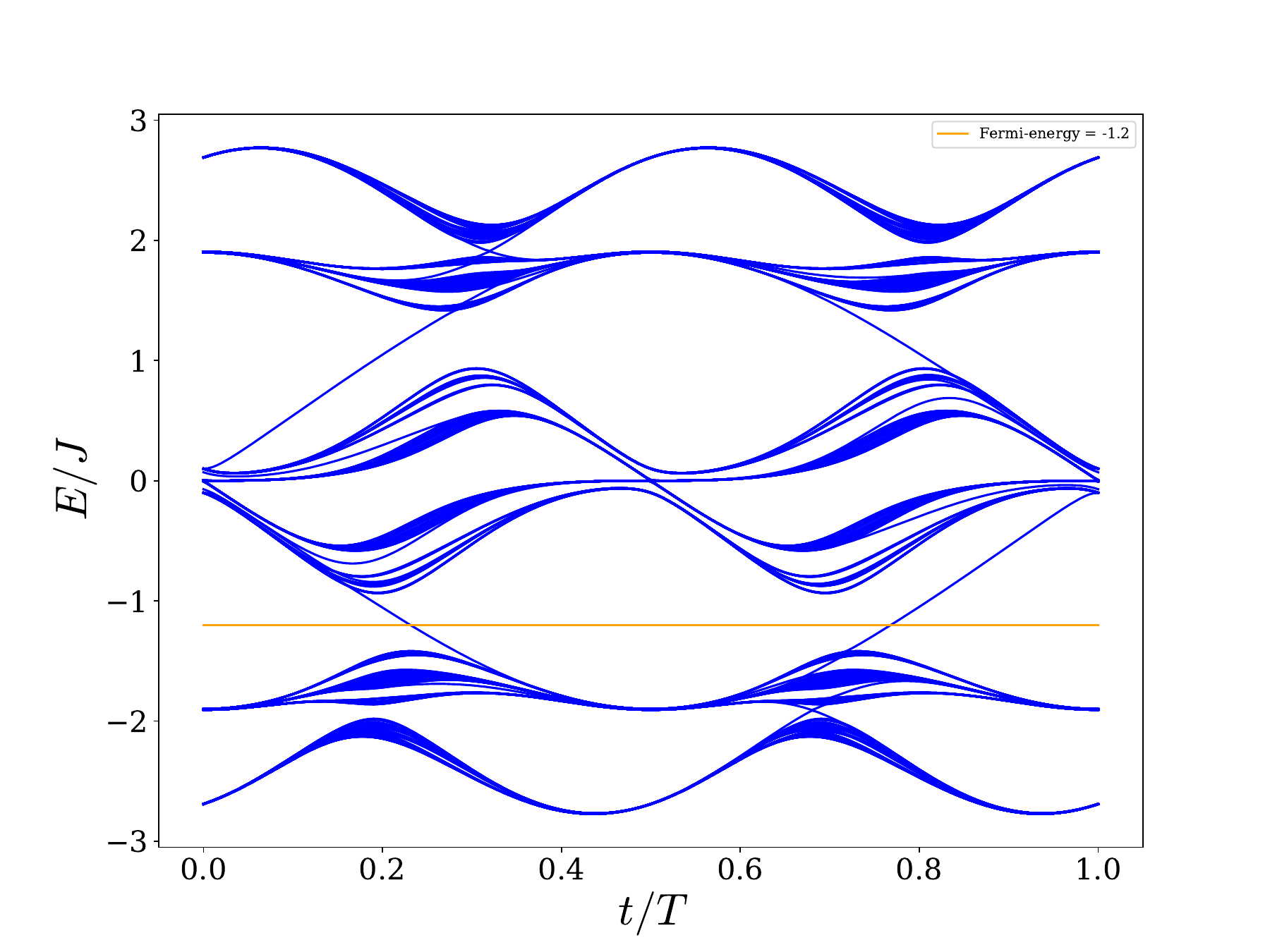}
    \caption{Energy spectrum of the Modified Rice-Mele model with open boundary conditions and the Thue-Morse sequence applied. The parameters of the model were set to $\delta_0 / J = 0.9$ and $\gamma_0 / J = 0.8$}
    \label{fig:Energy_spectrum_of_MRM_with_thue_morse_seq_J_1_delta0_point9_gamma0_point8}
\end{figure}

Previously we chose to apply the silver mean sequence, which is a quasiperiodic sequence, to the MRM model. It is then interesting to ask whether other aperiodic sequences that are not quasiperiodic also exhibit topological behavior. In this section we apply the Thue-Morse sequence as well as the period-doubling sequence to the MRM model to generate two aperiodic systems that do not have a quasicrystal nature. We then show that these non-quasicrystal aperidoic systems can demonstrate topological behavior.

We first consider the Thue-Morse aperiodic sequence applied to the MRM model. The Thue-Morse sequence is generated by setting the initial value of the sequence to $A$ and then using the following substitution rules; $g(A) = AB$ and $g(B) = BA$. This sequence is aperiodic but not quasiperiodic which is highlighted by the fact that the ratio of $A$ to $B$ for this sequence is one for any sequence length greater than one and not an irrational fraction like for the silver mean sequence or the Fibonacci sequence. Using \cref{binary conversion}, we apply the Thue-Morse sequence to the MRM and analyse its energy spectrum.

\Cref{fig:Energy_spectrum_of_MRM_with_thue_morse_seq_J_1_delta0_point9_gamma0_point8} shows the energy spectrum for the MRM with the Thue-Morse sequence applied and the parameters $\delta_0 / J = 0.9$ and $\gamma_0 / J = 0.8$. From this figure it can be seen that an energy gap exists at an energy of $-1.15$ and two edge state modes traverse this band gap, suggesting that it is topological in nature. Using $M_{1Q}(x,t)$ we confirmed that the gap is indeed topological in nature and has a topological index of one.

Next we consider the application of the period-doubling aperiodic sequence to the MRM model. The period-doubling sequence is generated with an initial value of $A$ and the substitution rules $g(A) = AB$ and $g(B) = AA$. This sequence, like the Thue-Morse sequence, is aperiodic but not quasiperiodic which is highlighted by the fact that the ratio of $A$ to $B$ for this sequence tends to the value of $2$ and not an irrational fraction. We again use \cref{binary conversion} to apply this sequence to the MRM model. 

\begin{figure}
    \includegraphics[width=0.48\textwidth, trim = {0, 2cm, 0, 2cm}]{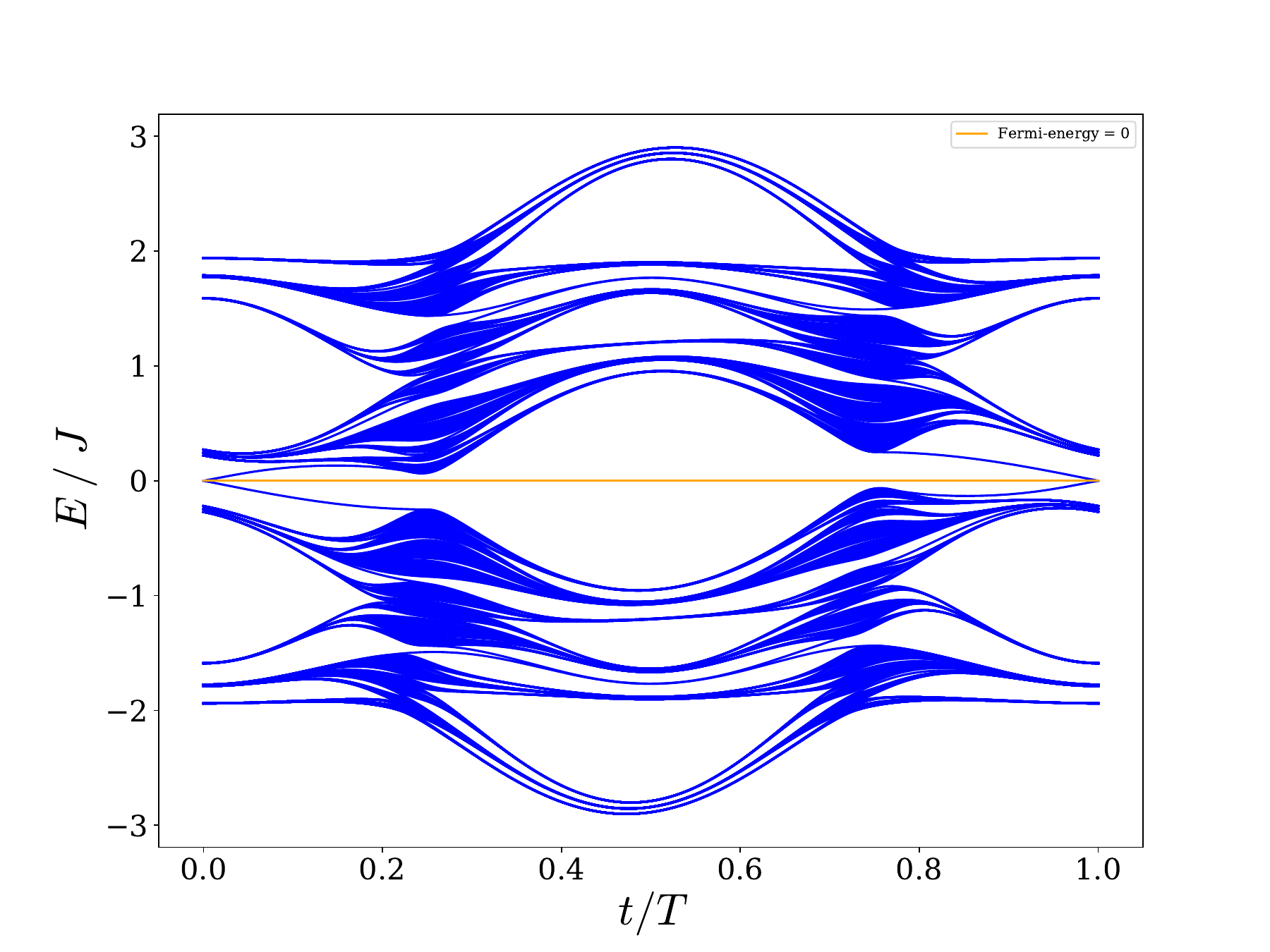}
    \caption{Energy spectrum of the Modified Rice-Mele model with open boundary conditions and the period-doubling sequence applied. The parameters of the model were set to $\delta_0 / J = 0.75$ and $\gamma_0 / J = 0.25$}
    \label{fig:Energy_spectrum_of_MRM_with_period_doubling_seq_J_1_delta0_point75_gamma0_point25}
\end{figure}

\Cref{fig:Energy_spectrum_of_MRM_with_period_doubling_seq_J_1_delta0_point75_gamma0_point25} shows the energy spectrum for the MRM with the period-doubling sequence. Here the parameters were set to $\delta_0 / J = 0.75$ and $\gamma_0 / J = 0.25$. This figure shows that for the period-doubling sequence an energy gap exists at an energy of $0$ and edge state modes traverse this band gap. Calculating the change in $M_{1Q}(x,t)$ we confirmed that the gap is topological in nature with a topological index of one. 

Both \cref{fig:Energy_spectrum_of_MRM_with_thue_morse_seq_J_1_delta0_point9_gamma0_point8} and \cref{fig:Energy_spectrum_of_MRM_with_period_doubling_seq_J_1_delta0_point75_gamma0_point25} show that topological properties can arise not just from quasicrystal ordering but also from general aperiodic ordering. As well as this, these figures show that, while the topological structure of the Thue-Morse chain and the period-doubling chain cannot currently be investigated directly \cite{zilberberg2021topology}, one can investigate the topological structure induced by these aperiodic sequences through the MRM model.

\section{quasicrystal 1D marker and Polarization} \label{polarization sec}

We have shown that the change in the 1D marker over a full time period can be used to determine the topological index of both periodic systems and aperiodic systems. However, it is also useful to know the polarization of the system and how it evolves over time. As such, in this section we show that the evolution of $M_{1Q}(x,t)$ over time quantitatively matches the evolution of the polarization of the system in the thermodynamic limit when $N_{\mathcal{R}} \to \infty$.

It is known that for a 1D system with open boundary conditions the polarization of the system is equal to the dipole moment per unit length which is given by \cite{resta1998quantum}
\begin{equation} \label{Tr_Px}
    P_{el}\,(t) = \frac{1}{N}{\rm Tr}(\hat{P}\hat{x}) = \frac{1}{N}\int dx \;x \;\rho(x)
\end{equation}
where the trace is over the whole of the system and $\rho(x)$ is the density. We will therefore compare the evolution of $M_{1Q}(x,t)$ to the evolution of \cref{Tr_Px} and show that they are equivalent when edge state effects are neglected.

\begin{figure}
    \includegraphics[width=0.48\textwidth, trim = {0, 2cm, 0, 2cm}]{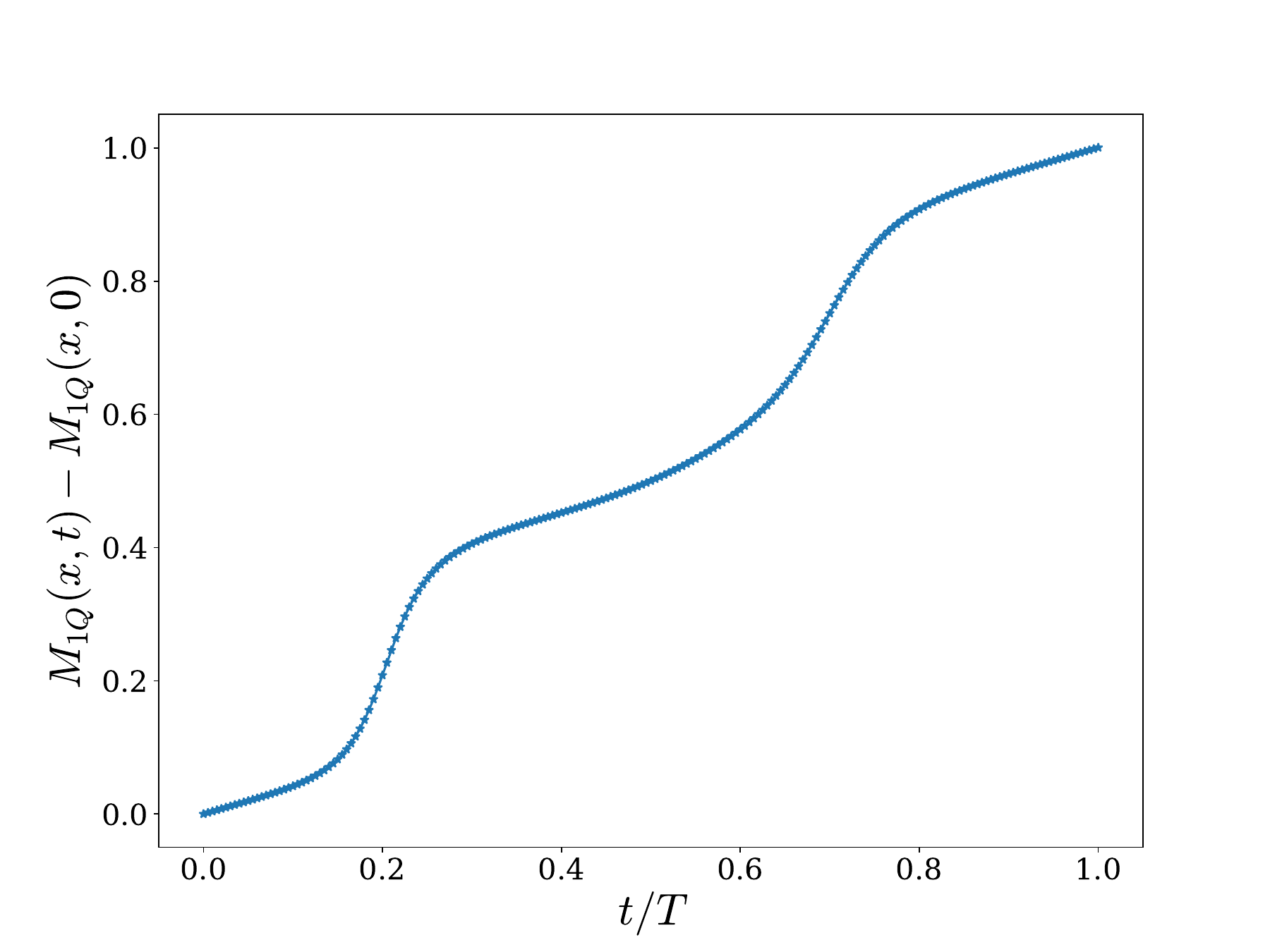}
    \caption{The time evolution of $M_{1Q}(x,t)$ over a full time period for the MRM with the silver mean sequence applied. The parameters of the system are the same as in \cref{fig:Change_in_M_1Q_for_varying_L_on_MRM_model_with_delta_0point9_gamma_0point8_FE_is_minus_1point45}. $N_{\mathcal{R}}$ spans all the lattice sites between, and including, $N=348$ and $N=1044$.}
    \label{fig:Change_in_M_1Q_for_MRM_model_with_delta_0point9_gamma_0point8_FE_is_minus_1point45}
\end{figure}

\Cref{fig:Change_in_M_1Q_for_MRM_model_with_delta_0point9_gamma_0point8_FE_is_minus_1point45} shows the evolution of $M_{1Q}(x,t)$ over a full time period for the MRM model with the silver mean sequence applied. The parameters of the system along with the Fermi-energy and the number of lattice sites were set to the values stated in \cref{fig:Change_in_M_1Q_for_varying_L_on_MRM_model_with_delta_0point9_gamma_0point8_FE_is_minus_1point45} and the region $\mathcal{R}$ encompassed the sites from $N=348$ to $N=1044$.  

\begin{figure}
    \includegraphics[width=0.48\textwidth, trim = {0, 2cm, 0, 2cm}]{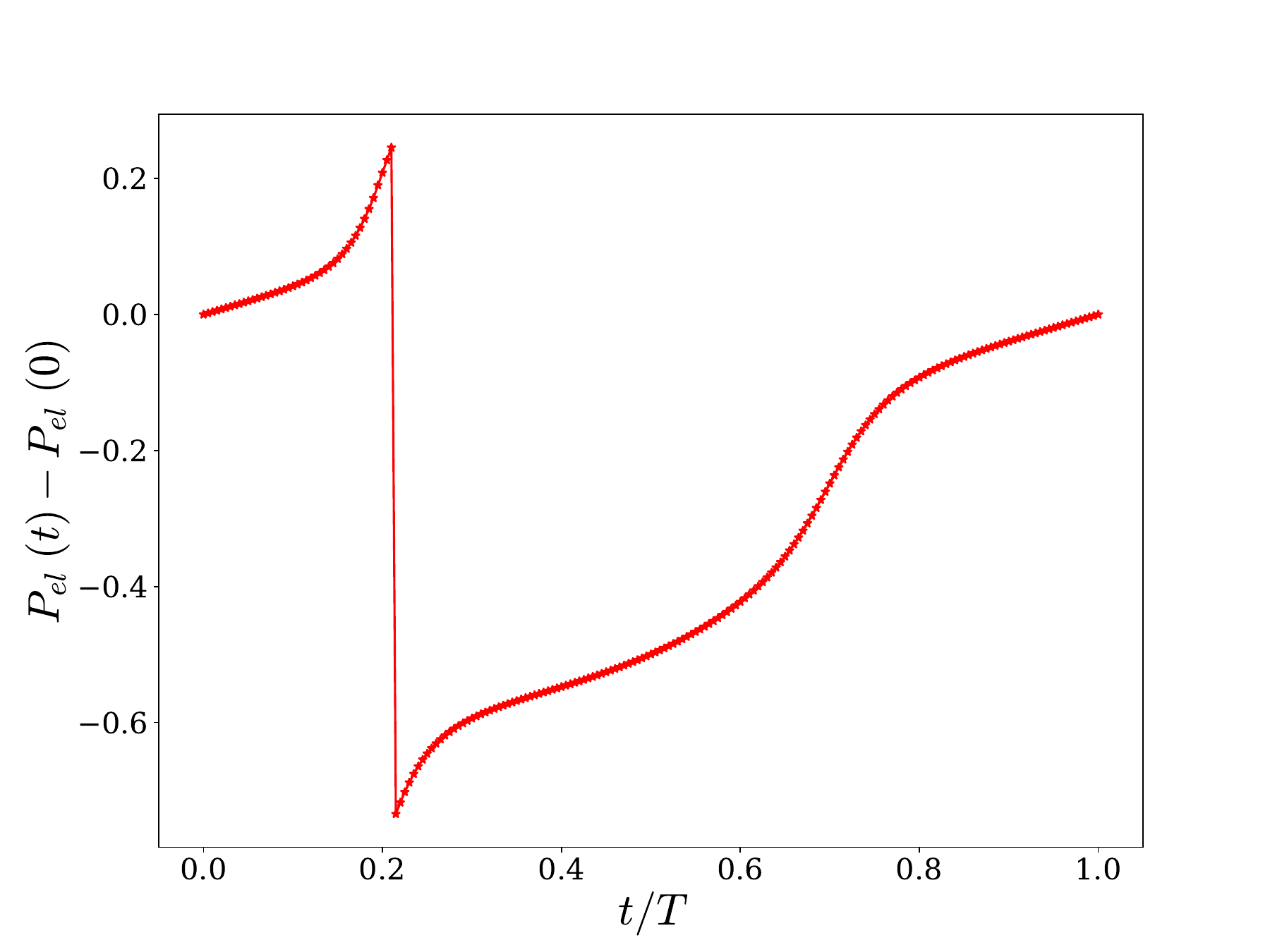}
    \caption{The time evolution of the polarization, $P_{el}$, over a full time period for the MRM with the silver mean sequence applied. The parameters of the system are the same as in \cref{fig:Change_in_M_1Q_for_varying_L_on_MRM_model_with_delta_0point9_gamma_0point8_FE_is_minus_1point45} and the polarization was calculated using \cref{Tr_Px}. The discontinuity is located at the point where the right edge state mode crosses the Fermi-energy causing a discontinuity in the polarization of magnitude one. The left edge state mode has a negligible effect on the polarization due to the fact we set the position of the far left site to 1.}
    \label{fig:Change_in_Tr_Px_for_MRM_model_with_delta_0point9_gamma_0point8_FE_is_minus_1point45}
\end{figure}

\cref{fig:Change_in_Tr_Px_for_MRM_model_with_delta_0point9_gamma_0point8_FE_is_minus_1point45} shows the polarization, $P_{el}\,(t)$, over a full time period for the same system. The discontinuity in $P_{el}\,(t)$ coincides with where the right edge state mode crosses the Fermi-energy and has a magnitude of approximately one. Such discontinuities of course will not occur for systems with periodic boundary conditions. On the other hand, using \cref{Tr_Px} to obtain the polarization is problematic for systems with periodic boundary conditions. The right edge state mode is exponentially localised at the position $N$ and using \cref{Tr_Px} it can be seen that its contribution to $P_{el}$ is approximately one. As such, when the right edge state mode crosses the Fermi-energy it is no longer considered in the polarization causing a discontinuity in $P_{el}$ of approximately one. This shows that the discontinuity is purely an edge state effect and seen as $M_{1Q}(x,t)$ only considers the bulk behavior we will ignore the discontinuity when comparing the evolution of the two. It should be noted that the left edge state mode does not cause a discontinuity in the polarization because we chose to label the far left site with position one meaning that the left edge state mode has a negligible contribution to the polarization of the system.

Comparing \cref{fig:Change_in_M_1Q_for_MRM_model_with_delta_0point9_gamma_0point8_FE_is_minus_1point45} and \cref{fig:Change_in_Tr_Px_for_MRM_model_with_delta_0point9_gamma_0point8_FE_is_minus_1point45} it can be shown that once the discontinuity is ignored the evolution of $M_{1Q}(x,t)$ quantitatively matches the evolution of $P_{el}\,(t)$. This then shows that the evolution of $M_{1Q}(x,t)$ can be used to describe the evolution of the bulk polarization of a system. It can also be shown that the discrepancy between the evolution of these two values decreases when $N_{\mathcal{R}}$ is increased and the edge state region $\mathcal{A}$, which contains all the lattice points not considered in region $\mathcal{R}$, is fixed. This then suggests that if one takes the thermodynamic limit by taking $N_{\mathcal{R}} \to \infty$ and keeping region $\mathcal{A}$ fixed and finite then the evolution of $M_{1Q}(x,t)$ tends to the evolution of the polarization of the system.

We have therefore shown that $M_{1Q}(x,t)$ can be used to accurately determine the evolution of the polarization over any period of time $t$ when the region $\mathcal{R}$ covers a large proportion of the system and in the thermodynamic limit the evolution of $M_{1Q}(x,t)$ tends to the evolution of the polarization. Though we have been primarily interested in topological systems in this work where the polarization changes by quantized values, it is important to note that this method works equally well for determining the bulk polarization of non-topological systems which is also of interest.

\section{Discussion and conclusion} \label{discus and conclus sec}

We have shown in this paper that the change in $M_{1Q}(x,t)$ can be used to determine the topological index of aperiodic systems where traditional methods of determining the topological index fail. We checked the topological index of the aperiodic system by analysing the flow of the Wannier centers across a partition in the bulk along with analysing the change in particle number on each side of this partition. It was also pointed out that one can consider the average behavior of a large group of Wannier centers in the bulk over a time period to gain a single value which can give the topological index of the system, similar to $M_{1Q}(x,t)$. However, we then showed that the accuracy of this method was less than that of $M_{1Q}(x,t)$. Lastly, we showed that the evolution of $M_{1Q}(x,t)$ quantitatively matches the evolution of the polarization of the system for large $N_{\mathcal{R}}$. Here we considered spinless models for ease, however, $M_{1Q}(x,t)$ can easily be applied to spin-dependent models in the same way.

Throughout this paper we have demonstrated that a system's topological index can be determined in a few different ways and therefore it is natural to ask why one should use the quasicrystal marker $M_{1Q}(x,t)$ over these other methods. While all these other methods have their place, the quasicrystal marker is a real space local topological marker that can vary over different regions of the system. This is useful when analysing systems with disorder and how this disorder affects the topological pumping through the system. The quasicrystal marker also produces a single valued output therefore making it desirable when analysing how the topological index changes with different parameters. Another important property of the quasicrystal marker is that it naturally generalizes to higher dimensional systems. If one uses the 3D generalization presented in \cite{sykes2021local} a 3D quasicrystal marker that determines the second Chern number of the system could easily be generated. Unlike the quasicrystal marker, a generalization to higher dimensions does not exist for the dipole moment or the Wannier center methods presented in this paper. For instance, as discussed in \cite{sykes2021local}, evaluating $\epsilon^{ijk}{\rm Tr}(\hat{P}\xh_i\hat{P}\xh_j\hat{P}\xh_k)$, where $\epsilon^{ijk}$ is the Levi-Civita symbol, for finite systems does not reveal topological behavior like that displayed for 1D in \cref{fig:Change_in_Tr_Px_for_MRM_model_with_delta_0point9_gamma_0point8_FE_is_minus_1point45}. We would also like to highlight that while we have used the quasicrystal marker on systems with open boundary conditions it can be used on systems with periodic boundary conditions as long as it is evaluated away from the region where the position operator, $\hat{x}$, is discontinuous which can be viewed as a ``branch cut" to be placed at a convenient point.

Another area in which topological physics is important is in non-Hermitian systems. It is therefore important to be able to identify the topological structure of these systems. As well as this, the bulk-boundary correspondence fails for certain non-Hermitian topological systems and, as such, a local topological marker could be useful for investigating these systems due to its local nature \cite{jin2019bulk}. Therefore, it would be useful if the 1D marker could be adapted so that it can be applied to these types of systems. It can be shown that the method used in this paper to evolve the projectors adiabatically in time can also be used for non-Hermitian Hamiltonians if one defines the projector as 
\begin{equation}
    \hat{P} = \sum_{n} \ket{\psi_n^R(t)}\bra{\psi_n^L(t)}.
\end{equation}
Here $\ket{\psi_n^R(t)}$ is the eigenvector of the Hamiltonian, $\hat{H}$, at time $t$, $\ket{\psi_n^L(t)}$ is the eigenvector of the Hermitian conjugate of the Hamiltonian, $\hat{H}^\dagger$, at time $t$ and $n$ labels the occupied bands. The eigenvectors are normalized such that $\braket{\psi_n^L}{\psi_{n'}^R} = \delta_{n,n'}$. Due to this, we expect that the 1D marker can be used to identify the topological structure of both periodic and aperiodic non-Hermitian systems when the projector is defined as above and $\hat{U}^\dagger$ is replaced with the inverse of the adiabatic evolution operator, $U^{-1}$; however, this needs to be confirmed.

It is important to point out that a time dependent version of the Bott index exists that has previously been used to analyse the topological index of an aperiodic system with periodic boundary conditions \cite{yoshii2021charge}. Whilst both the Bott index and $M_{QC}(x,t)$ can be used to determine the topological index of the system it is clear that these two objects differ in structure, the prime difference being the use of the adiabatic evolution operator in $M_{QC}(x,t)$. It would therefore be interesting to compare these two methods both analytically and computationally to see how they differ and what advantages each one possesses.

The models considered within this paper all smoothly evolve in time, which is not the case for all quasicrystal models. An example where this is not the case is the time dependent Fibonacci quasicrystal model where either the onsite potential or the hopping potential abruptly alternates between two values at some point in time \cite{levine1984quasicrystals, jagannathan2021fibonacci}. The abrupt change in the Hamiltonian means that adiabatic evolution is not possible for this model and consequently one cannot construct the adiabatic evolution operator meaning that $M_{1Q}(x,t)$ cannot correctly predict its topological index. From this we conclude that for $M_{1Q}(x,t)$ to correctly indicate the topological index of a system the system needs to be continuous in time to the degree that the adiabatic approximation can be applied. This is the only requirement needed to ensure that the change in the 1D quasicrystal marker accurately represents the topological index of a 1D time-dependent Hamiltonian.

\begin{acknowledgements}
We would like to thank 
Derek Lee, and
Peru d'Ornellas
for particularly useful discussions.  We gratefully acknowledge financial support from EPSRC Grant EP/N509486/1
and a Cecilia Tanner Research Impulse Grant from Imperial College Dept of Mathematics.
\end{acknowledgements}

\newpage

\providecommand{\noopsort}[1]{}\providecommand{\singleletter}[1]{#1}%


\begin{thebibliography}{39}%
\makeatletter
\providecommand \@ifxundefined [1]{%
 \@ifx{#1\undefined}
}%
\providecommand \@ifnum [1]{%
 \ifnum #1\expandafter \@firstoftwo
 \else \expandafter \@secondoftwo
 \fi
}%
\providecommand \@ifx [1]{%
 \ifx #1\expandafter \@firstoftwo
 \else \expandafter \@secondoftwo
 \fi
}%
\providecommand \natexlab [1]{#1}%
\providecommand \enquote  [1]{``#1''}%
\providecommand \bibnamefont  [1]{#1}%
\providecommand \bibfnamefont [1]{#1}%
\providecommand \citenamefont [1]{#1}%
\providecommand \href@noop [0]{\@secondoftwo}%
\providecommand \href [0]{\begingroup \@sanitize@url \@href}%
\providecommand \@href[1]{\@@startlink{#1}\@@href}%
\providecommand \@@href[1]{\endgroup#1\@@endlink}%
\providecommand \@sanitize@url [0]{\catcode `\\12\catcode `\$12\catcode
  `\&12\catcode `\#12\catcode `\^12\catcode `\_12\catcode `\%12\relax}%
\providecommand \@@startlink[1]{}%
\providecommand \@@endlink[0]{}%
\providecommand \url  [0]{\begingroup\@sanitize@url \@url }%
\providecommand \@url [1]{\endgroup\@href {#1}{\urlprefix }}%
\providecommand \urlprefix  [0]{URL }%
\providecommand \Eprint [0]{\href }%
\providecommand \doibase [0]{http://dx.doi.org/}%
\providecommand \selectlanguage [0]{\@gobble}%
\providecommand \bibinfo  [0]{\@secondoftwo}%
\providecommand \bibfield  [0]{\@secondoftwo}%
\providecommand \translation [1]{[#1]}%
\providecommand \BibitemOpen [0]{}%
\providecommand \bibitemStop [0]{}%
\providecommand \bibitemNoStop [0]{.\EOS\space}%
\providecommand \EOS [0]{\spacefactor3000\relax}%
\providecommand \BibitemShut  [1]{\csname bibitem#1\endcsname}%
\let\auto@bib@innerbib\@empty
\bibitem [{\citenamefont {Thouless}(1983)}]{thouless1983quantization}%
  \BibitemOpen
  \bibfield  {author} {\bibinfo {author} {\bibfnamefont {D.~J.}\ \bibnamefont
  {Thouless}},\ }\href@noop {} {\bibfield  {journal} {\bibinfo  {journal}
  {Phys. Rev. B}\ }\textbf {\bibinfo {volume} {27}},\ \bibinfo {pages} {6083}
  (\bibinfo {year} {1983})}\BibitemShut {NoStop}%
\bibitem [{\citenamefont {Thouless}\ \emph {et~al.}(1982)\citenamefont
  {Thouless}, \citenamefont {Kohmoto}, \citenamefont {Nightingale},\ and\
  \citenamefont {den Nijs}}]{TKNNpaper}%
  \BibitemOpen
  \bibfield  {author} {\bibinfo {author} {\bibfnamefont {D.~J.}\ \bibnamefont
  {Thouless}}, \bibinfo {author} {\bibfnamefont {M.}~\bibnamefont {Kohmoto}},
  \bibinfo {author} {\bibfnamefont {M.~P.}\ \bibnamefont {Nightingale}}, \ and\
  \bibinfo {author} {\bibfnamefont {M.}~\bibnamefont {den Nijs}},\ }\href
  {\doibase 10.1103/PhysRevLett.49.405} {\bibfield  {journal} {\bibinfo
  {journal} {Phys. Rev. Lett.}\ }\textbf {\bibinfo {volume} {49}},\ \bibinfo
  {pages} {405} (\bibinfo {year} {1982})}\BibitemShut {NoStop}%
\bibitem [{\citenamefont {Verbin}\ \emph {et~al.}(2015)\citenamefont {Verbin},
  \citenamefont {Zilberberg}, \citenamefont {Lahini}, \citenamefont {Kraus},\
  and\ \citenamefont {Silberberg}}]{verbin2015topological}%
  \BibitemOpen
  \bibfield  {author} {\bibinfo {author} {\bibfnamefont {M.}~\bibnamefont
  {Verbin}}, \bibinfo {author} {\bibfnamefont {O.}~\bibnamefont {Zilberberg}},
  \bibinfo {author} {\bibfnamefont {Y.}~\bibnamefont {Lahini}}, \bibinfo
  {author} {\bibfnamefont {Y.~E.}\ \bibnamefont {Kraus}}, \ and\ \bibinfo
  {author} {\bibfnamefont {Y.}~\bibnamefont {Silberberg}},\ }\href@noop {}
  {\bibfield  {journal} {\bibinfo  {journal} {Phys. Rev. B}\ }\textbf {\bibinfo
  {volume} {91}},\ \bibinfo {pages} {064201} (\bibinfo {year}
  {2015})}\BibitemShut {NoStop}%
\bibitem [{\citenamefont {Wimmer}\ \emph {et~al.}(2017)\citenamefont {Wimmer},
  \citenamefont {Price}, \citenamefont {Carusotto},\ and\ \citenamefont
  {Peschel}}]{wimmer2017experimental}%
  \BibitemOpen
  \bibfield  {author} {\bibinfo {author} {\bibfnamefont {M.}~\bibnamefont
  {Wimmer}}, \bibinfo {author} {\bibfnamefont {H.~M.}\ \bibnamefont {Price}},
  \bibinfo {author} {\bibfnamefont {I.}~\bibnamefont {Carusotto}}, \ and\
  \bibinfo {author} {\bibfnamefont {U.}~\bibnamefont {Peschel}},\ }\href@noop
  {} {\bibfield  {journal} {\bibinfo  {journal} {Nat. Phys.}\ }\textbf
  {\bibinfo {volume} {13}},\ \bibinfo {pages} {545} (\bibinfo {year}
  {2017})}\BibitemShut {NoStop}%
\bibitem [{\citenamefont {Lohse}\ \emph {et~al.}(2016)\citenamefont {Lohse},
  \citenamefont {Schweizer}, \citenamefont {Zilberberg}, \citenamefont
  {Aidelsburger},\ and\ \citenamefont {Bloch}}]{lohse2016thouless}%
  \BibitemOpen
  \bibfield  {author} {\bibinfo {author} {\bibfnamefont {M.}~\bibnamefont
  {Lohse}}, \bibinfo {author} {\bibfnamefont {C.}~\bibnamefont {Schweizer}},
  \bibinfo {author} {\bibfnamefont {O.}~\bibnamefont {Zilberberg}}, \bibinfo
  {author} {\bibfnamefont {M.}~\bibnamefont {Aidelsburger}}, \ and\ \bibinfo
  {author} {\bibfnamefont {I.}~\bibnamefont {Bloch}},\ }\href@noop {}
  {\bibfield  {journal} {\bibinfo  {journal} {Nat. Phys.}\ }\textbf {\bibinfo
  {volume} {12}},\ \bibinfo {pages} {350} (\bibinfo {year} {2016})}\BibitemShut
  {NoStop}%
\bibitem [{\citenamefont {Schweizer}\ \emph {et~al.}(2016)\citenamefont
  {Schweizer}, \citenamefont {Lohse}, \citenamefont {Citro},\ and\
  \citenamefont {Bloch}}]{schweizer2016spin}%
  \BibitemOpen
  \bibfield  {author} {\bibinfo {author} {\bibfnamefont {C.}~\bibnamefont
  {Schweizer}}, \bibinfo {author} {\bibfnamefont {M.}~\bibnamefont {Lohse}},
  \bibinfo {author} {\bibfnamefont {R.}~\bibnamefont {Citro}}, \ and\ \bibinfo
  {author} {\bibfnamefont {I.}~\bibnamefont {Bloch}},\ }\href@noop {}
  {\bibfield  {journal} {\bibinfo  {journal} {Phys. Rev. Lett.}\ }\textbf
  {\bibinfo {volume} {117}},\ \bibinfo {pages} {170405} (\bibinfo {year}
  {2016})}\BibitemShut {NoStop}%
\bibitem [{\citenamefont {Nakajima}\ \emph {et~al.}(2016)\citenamefont
  {Nakajima}, \citenamefont {Tomita}, \citenamefont {Taie}, \citenamefont
  {Ichinose}, \citenamefont {Ozawa}, \citenamefont {Wang}, \citenamefont
  {Troyer},\ and\ \citenamefont {Takahashi}}]{nakajima2016topological}%
  \BibitemOpen
  \bibfield  {author} {\bibinfo {author} {\bibfnamefont {S.}~\bibnamefont
  {Nakajima}}, \bibinfo {author} {\bibfnamefont {T.}~\bibnamefont {Tomita}},
  \bibinfo {author} {\bibfnamefont {S.}~\bibnamefont {Taie}}, \bibinfo {author}
  {\bibfnamefont {T.}~\bibnamefont {Ichinose}}, \bibinfo {author}
  {\bibfnamefont {H.}~\bibnamefont {Ozawa}}, \bibinfo {author} {\bibfnamefont
  {L.}~\bibnamefont {Wang}}, \bibinfo {author} {\bibfnamefont {M.}~\bibnamefont
  {Troyer}}, \ and\ \bibinfo {author} {\bibfnamefont {Y.}~\bibnamefont
  {Takahashi}},\ }\href@noop {} {\bibfield  {journal} {\bibinfo  {journal}
  {Nat. Phys.}\ }\textbf {\bibinfo {volume} {12}},\ \bibinfo {pages} {296}
  (\bibinfo {year} {2016})}\BibitemShut {NoStop}%
\bibitem [{\citenamefont {Nakajima}\ \emph {et~al.}(2021)\citenamefont
  {Nakajima}, \citenamefont {Takei}, \citenamefont {Sakuma}, \citenamefont
  {Kuno}, \citenamefont {Marra},\ and\ \citenamefont
  {Takahashi}}]{nakajima2021competition}%
  \BibitemOpen
  \bibfield  {author} {\bibinfo {author} {\bibfnamefont {S.}~\bibnamefont
  {Nakajima}}, \bibinfo {author} {\bibfnamefont {N.}~\bibnamefont {Takei}},
  \bibinfo {author} {\bibfnamefont {K.}~\bibnamefont {Sakuma}}, \bibinfo
  {author} {\bibfnamefont {Y.}~\bibnamefont {Kuno}}, \bibinfo {author}
  {\bibfnamefont {P.}~\bibnamefont {Marra}}, \ and\ \bibinfo {author}
  {\bibfnamefont {Y.}~\bibnamefont {Takahashi}},\ }\href@noop {} {\bibfield
  {journal} {\bibinfo  {journal} {Nature Physics}\ ,\ \bibinfo {pages} {1}}
  (\bibinfo {year} {2021})}\BibitemShut {NoStop}%
\bibitem [{\citenamefont {Kraus}\ and\ \citenamefont
  {Zilberberg}(2012)}]{kraus2012equivalence}%
  \BibitemOpen
  \bibfield  {author} {\bibinfo {author} {\bibfnamefont {Y.~E.}\ \bibnamefont
  {Kraus}}\ and\ \bibinfo {author} {\bibfnamefont {O.}~\bibnamefont
  {Zilberberg}},\ }\href@noop {} {\bibfield  {journal} {\bibinfo  {journal}
  {Physical review letters}\ }\textbf {\bibinfo {volume} {109}},\ \bibinfo
  {pages} {116404} (\bibinfo {year} {2012})}\BibitemShut {NoStop}%
\bibitem [{\citenamefont {Zilberberg}(2021)}]{zilberberg2021topology}%
  \BibitemOpen
  \bibfield  {author} {\bibinfo {author} {\bibfnamefont {O.}~\bibnamefont
  {Zilberberg}},\ }\href@noop {} {\bibfield  {journal} {\bibinfo  {journal}
  {Optical Materials Express}\ }\textbf {\bibinfo {volume} {11}},\ \bibinfo
  {pages} {1143} (\bibinfo {year} {2021})}\BibitemShut {NoStop}%
\bibitem [{\citenamefont {Jagannathan}(2021)}]{jagannathan2021fibonacci}%
  \BibitemOpen
  \bibfield  {author} {\bibinfo {author} {\bibfnamefont {A.}~\bibnamefont
  {Jagannathan}},\ }\href@noop {} {\bibfield  {journal} {\bibinfo  {journal}
  {Reviews of Modern Physics}\ }\textbf {\bibinfo {volume} {93}},\ \bibinfo
  {pages} {045001} (\bibinfo {year} {2021})}\BibitemShut {NoStop}%
\bibitem [{\citenamefont {Aubry}\ and\ \citenamefont
  {Andr{\'e}}(1980)}]{aubry1980analyticity}%
  \BibitemOpen
  \bibfield  {author} {\bibinfo {author} {\bibfnamefont {S.}~\bibnamefont
  {Aubry}}\ and\ \bibinfo {author} {\bibfnamefont {G.}~\bibnamefont
  {Andr{\'e}}},\ }\href@noop {} {\bibfield  {journal} {\bibinfo  {journal}
  {Ann. Israel. Phys. Soc}\ }\textbf {\bibinfo {volume} {3}},\ \bibinfo {pages}
  {133} (\bibinfo {year} {1980})}\BibitemShut {NoStop}%
\bibitem [{\citenamefont {Bianco}\ and\ \citenamefont
  {Resta}(2011)}]{bianco2011mapping}%
  \BibitemOpen
  \bibfield  {author} {\bibinfo {author} {\bibfnamefont {R.}~\bibnamefont
  {Bianco}}\ and\ \bibinfo {author} {\bibfnamefont {R.}~\bibnamefont {Resta}},\
  }\href@noop {} {\bibfield  {journal} {\bibinfo  {journal} {Phys. Rev. B}\
  }\textbf {\bibinfo {volume} {84}},\ \bibinfo {pages} {241106(R)} (\bibinfo
  {year} {2011})}\BibitemShut {NoStop}%
\bibitem [{\citenamefont {Prodan}\ \emph {et~al.}(2010)\citenamefont {Prodan},
  \citenamefont {Hughes},\ and\ \citenamefont
  {Bernevig}}]{Prodan2010Entanglement}%
  \BibitemOpen
  \bibfield  {author} {\bibinfo {author} {\bibfnamefont {E.}~\bibnamefont
  {Prodan}}, \bibinfo {author} {\bibfnamefont {T.~L.}\ \bibnamefont {Hughes}},
  \ and\ \bibinfo {author} {\bibfnamefont {B.~A.}\ \bibnamefont {Bernevig}},\
  }\href {\doibase 10.1103/PhysRevLett.105.115501} {\bibfield  {journal}
  {\bibinfo  {journal} {Phys. Rev. Lett.}\ }\textbf {\bibinfo {volume} {105}},\
  \bibinfo {pages} {115501} (\bibinfo {year} {2010})}\BibitemShut {NoStop}%
\bibitem [{\citenamefont {Tran}\ \emph {et~al.}(2015)\citenamefont {Tran},
  \citenamefont {Dauphin}, \citenamefont {Goldman},\ and\ \citenamefont
  {Gaspard}}]{tran2015topological}%
  \BibitemOpen
  \bibfield  {author} {\bibinfo {author} {\bibfnamefont {D.~T.}\ \bibnamefont
  {Tran}}, \bibinfo {author} {\bibfnamefont {A.}~\bibnamefont {Dauphin}},
  \bibinfo {author} {\bibfnamefont {N.}~\bibnamefont {Goldman}}, \ and\
  \bibinfo {author} {\bibfnamefont {P.}~\bibnamefont {Gaspard}},\ }\href@noop
  {} {\bibfield  {journal} {\bibinfo  {journal} {Phys. Rev. B}\ }\textbf
  {\bibinfo {volume} {91}},\ \bibinfo {pages} {085125} (\bibinfo {year}
  {2015})}\BibitemShut {NoStop}%
\bibitem [{\citenamefont {Marrazzo}\ and\ \citenamefont
  {Resta}(2017)}]{marrazzo2017locality}%
  \BibitemOpen
  \bibfield  {author} {\bibinfo {author} {\bibfnamefont {A.}~\bibnamefont
  {Marrazzo}}\ and\ \bibinfo {author} {\bibfnamefont {R.}~\bibnamefont
  {Resta}},\ }\href@noop {} {\bibfield  {journal} {\bibinfo  {journal} {Phys.
  Rev. B}\ }\textbf {\bibinfo {volume} {95}},\ \bibinfo {pages} {121114(R)}
  (\bibinfo {year} {2017})}\BibitemShut {NoStop}%
\bibitem [{\citenamefont {Mitchell}\ \emph {et~al.}(2018)\citenamefont
  {Mitchell}, \citenamefont {Nash}, \citenamefont {Hexner}, \citenamefont
  {Turner},\ and\ \citenamefont {Irvine}}]{mitchell2018amorphous}%
  \BibitemOpen
  \bibfield  {author} {\bibinfo {author} {\bibfnamefont {N.~P.}\ \bibnamefont
  {Mitchell}}, \bibinfo {author} {\bibfnamefont {L.~M.}\ \bibnamefont {Nash}},
  \bibinfo {author} {\bibfnamefont {D.}~\bibnamefont {Hexner}}, \bibinfo
  {author} {\bibfnamefont {A.~M.}\ \bibnamefont {Turner}}, \ and\ \bibinfo
  {author} {\bibfnamefont {W.~T.}\ \bibnamefont {Irvine}},\ }\href@noop {}
  {\bibfield  {journal} {\bibinfo  {journal} {Nat. Phys.}\ }\textbf {\bibinfo
  {volume} {14}},\ \bibinfo {pages} {380} (\bibinfo {year} {2018})}\BibitemShut
  {NoStop}%
\bibitem [{\citenamefont {Irsigler}\ \emph {et~al.}(2019)\citenamefont
  {Irsigler}, \citenamefont {Zheng},\ and\ \citenamefont
  {Hofstetter}}]{Irsigler2019Interacting}%
  \BibitemOpen
  \bibfield  {author} {\bibinfo {author} {\bibfnamefont {B.}~\bibnamefont
  {Irsigler}}, \bibinfo {author} {\bibfnamefont {J.~H.}\ \bibnamefont {Zheng}},
  \ and\ \bibinfo {author} {\bibfnamefont {W.}~\bibnamefont {Hofstetter}},\
  }\href {\doibase 10.1103/PhysRevLett.122.010406} {\bibfield  {journal}
  {\bibinfo  {journal} {Phys. Rev. Lett.}\ }\textbf {\bibinfo {volume} {122}},\
  \bibinfo {pages} {010406} (\bibinfo {year} {2019})}\BibitemShut {NoStop}%
\bibitem [{\citenamefont {Gebert}\ \emph {et~al.}(2020)\citenamefont {Gebert},
  \citenamefont {Irsigler},\ and\ \citenamefont
  {Hofstetter}}]{Gerbert2020Local}%
  \BibitemOpen
  \bibfield  {author} {\bibinfo {author} {\bibfnamefont {U.}~\bibnamefont
  {Gebert}}, \bibinfo {author} {\bibfnamefont {B.}~\bibnamefont {Irsigler}}, \
  and\ \bibinfo {author} {\bibfnamefont {W.}~\bibnamefont {Hofstetter}},\
  }\href {\doibase 10.1103/PhysRevA.101.063606} {\bibfield  {journal} {\bibinfo
   {journal} {Phys. Rev. A}\ }\textbf {\bibinfo {volume} {101}},\ \bibinfo
  {pages} {063606} (\bibinfo {year} {2020})}\BibitemShut {NoStop}%
\bibitem [{\citenamefont {Ul\ifmmode~\check{c}\else \v{c}\fi{}akar}\ \emph
  {et~al.}(2020)\citenamefont {Ul\ifmmode~\check{c}\else \v{c}\fi{}akar},
  \citenamefont {Mravlje},\ and\ \citenamefont {Rejec}}]{ulcakar2020kibble}%
  \BibitemOpen
  \bibfield  {author} {\bibinfo {author} {\bibfnamefont {L.}~\bibnamefont
  {Ul\ifmmode~\check{c}\else \v{c}\fi{}akar}}, \bibinfo {author} {\bibfnamefont
  {J.}~\bibnamefont {Mravlje}}, \ and\ \bibinfo {author} {\bibfnamefont
  {T.~c.~v.}\ \bibnamefont {Rejec}},\ }\href {\doibase
  10.1103/PhysRevLett.125.216601} {\bibfield  {journal} {\bibinfo  {journal}
  {Phys. Rev. Lett.}\ }\textbf {\bibinfo {volume} {125}},\ \bibinfo {pages}
  {216601} (\bibinfo {year} {2020})}\BibitemShut {NoStop}%
\bibitem [{\citenamefont {Hayward}\ \emph {et~al.}(2021)\citenamefont
  {Hayward}, \citenamefont {Bertok}, \citenamefont {Schneider},\ and\
  \citenamefont {Heidrich-Meisner}}]{hayward2020effect}%
  \BibitemOpen
  \bibfield  {author} {\bibinfo {author} {\bibfnamefont {A.~L.~C.}\
  \bibnamefont {Hayward}}, \bibinfo {author} {\bibfnamefont {E.}~\bibnamefont
  {Bertok}}, \bibinfo {author} {\bibfnamefont {U.}~\bibnamefont {Schneider}}, \
  and\ \bibinfo {author} {\bibfnamefont {F.}~\bibnamefont {Heidrich-Meisner}},\
  }\href {\doibase 10.1103/PhysRevA.103.043310} {\bibfield  {journal} {\bibinfo
   {journal} {Phys. Rev. A}\ }\textbf {\bibinfo {volume} {103}},\ \bibinfo
  {pages} {043310} (\bibinfo {year} {2021})}\BibitemShut {NoStop}%
\bibitem [{\citenamefont {Varjas}\ \emph {et~al.}(2020)\citenamefont {Varjas},
  \citenamefont {Fruchart}, \citenamefont {Akhmerov},\ and\ \citenamefont
  {Perez-Piskunow}}]{varjas2020computation}%
  \BibitemOpen
  \bibfield  {author} {\bibinfo {author} {\bibfnamefont {D.}~\bibnamefont
  {Varjas}}, \bibinfo {author} {\bibfnamefont {M.}~\bibnamefont {Fruchart}},
  \bibinfo {author} {\bibfnamefont {A.~R.}\ \bibnamefont {Akhmerov}}, \ and\
  \bibinfo {author} {\bibfnamefont {P.~M.}\ \bibnamefont {Perez-Piskunow}},\
  }\href@noop {} {\bibfield  {journal} {\bibinfo  {journal} {Phys. Rev.
  Research}\ }\textbf {\bibinfo {volume} {2}},\ \bibinfo {pages} {013229}
  (\bibinfo {year} {2020})}\BibitemShut {NoStop}%
\bibitem [{\citenamefont {Ghadimi}\ \emph {et~al.}(2021)\citenamefont
  {Ghadimi}, \citenamefont {Sugimoto}, \citenamefont {Tanaka},\ and\
  \citenamefont {Tohyama}}]{ghadimi2021topological}%
  \BibitemOpen
  \bibfield  {author} {\bibinfo {author} {\bibfnamefont {R.}~\bibnamefont
  {Ghadimi}}, \bibinfo {author} {\bibfnamefont {T.}~\bibnamefont {Sugimoto}},
  \bibinfo {author} {\bibfnamefont {K.}~\bibnamefont {Tanaka}}, \ and\ \bibinfo
  {author} {\bibfnamefont {T.}~\bibnamefont {Tohyama}},\ }\href@noop {}
  {\bibfield  {journal} {\bibinfo  {journal} {Physical Review B}\ }\textbf
  {\bibinfo {volume} {104}},\ \bibinfo {pages} {144511} (\bibinfo {year}
  {2021})}\BibitemShut {NoStop}%
\bibitem [{\citenamefont {Johnstone}\ \emph {et~al.}()\citenamefont
  {Johnstone}, \citenamefont {Colbrook}, \citenamefont {Nielsen}, \citenamefont
  {{\"O}hberg},\ and\ \citenamefont {Duncan}}]{johnstone2021bulk}%
  \BibitemOpen
  \bibfield  {author} {\bibinfo {author} {\bibfnamefont {D.}~\bibnamefont
  {Johnstone}}, \bibinfo {author} {\bibfnamefont {M.~J.}\ \bibnamefont
  {Colbrook}}, \bibinfo {author} {\bibfnamefont {A.~E.}\ \bibnamefont
  {Nielsen}}, \bibinfo {author} {\bibfnamefont {P.}~\bibnamefont {{\"O}hberg}},
  \ and\ \bibinfo {author} {\bibfnamefont {C.~W.}\ \bibnamefont {Duncan}},\
  }\href@noop {} {\bibinfo  {journal} {arXiv:2107.05635}\ }\BibitemShut
  {NoStop}%
\bibitem [{\citenamefont {Sykes}\ and\ \citenamefont
  {Barnett}(2021)}]{sykes2021local}%
  \BibitemOpen
\bibfield  {journal} {  }\bibfield  {author} {\bibinfo {author} {\bibfnamefont
  {J.}~\bibnamefont {Sykes}}\ and\ \bibinfo {author} {\bibfnamefont
  {R.}~\bibnamefont {Barnett}},\ }\href@noop {} {\bibfield  {journal} {\bibinfo
   {journal} {Physical Review B}\ }\textbf {\bibinfo {volume} {103}},\ \bibinfo
  {pages} {155134} (\bibinfo {year} {2021})}\BibitemShut {NoStop}%
\bibitem [{\citenamefont {Yoshii}\ \emph {et~al.}(2021)\citenamefont {Yoshii},
  \citenamefont {Kitamura},\ and\ \citenamefont {Morimoto}}]{yoshii2021charge}%
  \BibitemOpen
  \bibfield  {author} {\bibinfo {author} {\bibfnamefont {M.}~\bibnamefont
  {Yoshii}}, \bibinfo {author} {\bibfnamefont {S.}~\bibnamefont {Kitamura}}, \
  and\ \bibinfo {author} {\bibfnamefont {T.}~\bibnamefont {Morimoto}},\ }\href
  {\doibase 10.1103/PhysRevB.104.155126} {\bibfield  {journal} {\bibinfo
  {journal} {Phys. Rev. B}\ }\textbf {\bibinfo {volume} {104}},\ \bibinfo
  {pages} {155126} (\bibinfo {year} {2021})}\BibitemShut {NoStop}%
\bibitem [{\citenamefont {Kato}(1950)}]{kato1950adiabatic}%
  \BibitemOpen
  \bibfield  {author} {\bibinfo {author} {\bibfnamefont {T.}~\bibnamefont
  {Kato}},\ }\href@noop {} {\bibfield  {journal} {\bibinfo  {journal} {J. Phys.
  Soc. Japan}\ }\textbf {\bibinfo {volume} {5}},\ \bibinfo {pages} {435}
  (\bibinfo {year} {1950})}\BibitemShut {NoStop}%
\bibitem [{\citenamefont {Messiah}(1981)}]{messiah1981quantum}%
  \BibitemOpen
  \bibfield  {author} {\bibinfo {author} {\bibfnamefont {A.}~\bibnamefont
  {Messiah}},\ }\href@noop {} {\emph {\bibinfo {title} {Quantum mechanics}}},\
  Vol.~\bibinfo {volume} {2}\ (\bibinfo  {publisher} {Elsevier},\ \bibinfo
  {year} {1981})\BibitemShut {NoStop}%
\bibitem [{\citenamefont {Kitaev}(2006)}]{kitaev2006anyons}%
  \BibitemOpen
  \bibfield  {author} {\bibinfo {author} {\bibfnamefont {A.}~\bibnamefont
  {Kitaev}},\ }\href@noop {} {\bibfield  {journal} {\bibinfo  {journal} {Ann.
  Phys. (N. Y.)}\ }\textbf {\bibinfo {volume} {321}},\ \bibinfo {pages} {2}
  (\bibinfo {year} {2006})}\BibitemShut {NoStop}%
\bibitem [{\citenamefont {Bellissard}\ \emph {et~al.}(1989)\citenamefont
  {Bellissard}, \citenamefont {Iochum}, \citenamefont {Scoppola},\ and\
  \citenamefont {Testard}}]{bellissard1989spectral}%
  \BibitemOpen
  \bibfield  {author} {\bibinfo {author} {\bibfnamefont {J.}~\bibnamefont
  {Bellissard}}, \bibinfo {author} {\bibfnamefont {B.}~\bibnamefont {Iochum}},
  \bibinfo {author} {\bibfnamefont {E.}~\bibnamefont {Scoppola}}, \ and\
  \bibinfo {author} {\bibfnamefont {D.}~\bibnamefont {Testard}},\ }\href@noop
  {} {\bibfield  {journal} {\bibinfo  {journal} {Communications in Mathematical
  Physics}\ }\textbf {\bibinfo {volume} {125}},\ \bibinfo {pages} {527}
  (\bibinfo {year} {1989})}\BibitemShut {NoStop}%
\bibitem [{\citenamefont {Kalish}\ \emph {et~al.}(2018)\citenamefont {Kalish},
  \citenamefont {Komarov}, \citenamefont {Kozhaev}, \citenamefont {Achanta},
  \citenamefont {Dagesyan}, \citenamefont {Shaposhnikov}, \citenamefont
  {Prokopov}, \citenamefont {Berzhansky}, \citenamefont {Zvezdin},\ and\
  \citenamefont {Belotelov}}]{kalish2018magnetoplasmonic}%
  \BibitemOpen
  \bibfield  {author} {\bibinfo {author} {\bibfnamefont {A.~N.}\ \bibnamefont
  {Kalish}}, \bibinfo {author} {\bibfnamefont {R.~S.}\ \bibnamefont {Komarov}},
  \bibinfo {author} {\bibfnamefont {M.~A.}\ \bibnamefont {Kozhaev}}, \bibinfo
  {author} {\bibfnamefont {V.~G.}\ \bibnamefont {Achanta}}, \bibinfo {author}
  {\bibfnamefont {S.~A.}\ \bibnamefont {Dagesyan}}, \bibinfo {author}
  {\bibfnamefont {A.~N.}\ \bibnamefont {Shaposhnikov}}, \bibinfo {author}
  {\bibfnamefont {A.~R.}\ \bibnamefont {Prokopov}}, \bibinfo {author}
  {\bibfnamefont {V.~N.}\ \bibnamefont {Berzhansky}}, \bibinfo {author}
  {\bibfnamefont {A.~K.}\ \bibnamefont {Zvezdin}}, \ and\ \bibinfo {author}
  {\bibfnamefont {V.~I.}\ \bibnamefont {Belotelov}},\ }\href@noop {} {\bibfield
   {journal} {\bibinfo  {journal} {Optica}\ }\textbf {\bibinfo {volume} {5}},\
  \bibinfo {pages} {617} (\bibinfo {year} {2018})}\BibitemShut {NoStop}%
\bibitem [{\citenamefont {Baake}\ and\ \citenamefont
  {Grimm}(2013)}]{baake_grimm_2013}%
  \BibitemOpen
  \bibfield  {author} {\bibinfo {author} {\bibfnamefont {M.}~\bibnamefont
  {Baake}}\ and\ \bibinfo {author} {\bibfnamefont {U.}~\bibnamefont {Grimm}},\
  }\href {\doibase 10.1017/CBO9781139025256} {\emph {\bibinfo {title}
  {Aperiodic Order}}},\ \bibinfo {series} {Encyclopedia of Mathematics and its
  Applications}, Vol.~\bibinfo {volume} {1}\ (\bibinfo  {publisher} {Cambridge
  University Press},\ \bibinfo {year} {2013})\BibitemShut {NoStop}%
\bibitem [{\citenamefont {Asb{\'o}th}\ \emph {et~al.}(2016)\citenamefont
  {Asb{\'o}th}, \citenamefont {Oroszl{\'a}ny},\ and\ \citenamefont
  {P{\'a}lyi}}]{asboth2016short}%
  \BibitemOpen
  \bibfield  {author} {\bibinfo {author} {\bibfnamefont {J.~K.}\ \bibnamefont
  {Asb{\'o}th}}, \bibinfo {author} {\bibfnamefont {L.}~\bibnamefont
  {Oroszl{\'a}ny}}, \ and\ \bibinfo {author} {\bibfnamefont {A.}~\bibnamefont
  {P{\'a}lyi}},\ }\href {\doibase 10.1007/9783319256078} {\emph {\bibinfo
  {title} {A short course on topological insulators}}}\ (\bibinfo  {publisher}
  {Springer},\ \bibinfo {year} {2016})\BibitemShut {NoStop}%
\bibitem [{\citenamefont {Kivelson}(1982)}]{kivelson1982wannier}%
  \BibitemOpen
  \bibfield  {author} {\bibinfo {author} {\bibfnamefont {S.}~\bibnamefont
  {Kivelson}},\ }\href@noop {} {\bibfield  {journal} {\bibinfo  {journal}
  {Physical Review B}\ }\textbf {\bibinfo {volume} {26}},\ \bibinfo {pages}
  {4269} (\bibinfo {year} {1982})}\BibitemShut {NoStop}%
\bibitem [{\citenamefont {Niu}(1991)}]{niu1991theory}%
  \BibitemOpen
  \bibfield  {author} {\bibinfo {author} {\bibfnamefont {Q.}~\bibnamefont
  {Niu}},\ }\href@noop {} {\bibfield  {journal} {\bibinfo  {journal} {Modern
  Physics Letters B}\ }\textbf {\bibinfo {volume} {5}},\ \bibinfo {pages} {923}
  (\bibinfo {year} {1991})}\BibitemShut {NoStop}%
\bibitem [{\citenamefont {Nenciu}\ and\ \citenamefont
  {Nenciu}(1998)}]{nenciu1998existence}%
  \BibitemOpen
  \bibfield  {author} {\bibinfo {author} {\bibfnamefont {A.}~\bibnamefont
  {Nenciu}}\ and\ \bibinfo {author} {\bibfnamefont {G.}~\bibnamefont
  {Nenciu}},\ }\href@noop {} {\bibfield  {journal} {\bibinfo  {journal}
  {Communications in mathematical physics}\ }\textbf {\bibinfo {volume}
  {190}},\ \bibinfo {pages} {541} (\bibinfo {year} {1998})}\BibitemShut
  {NoStop}%
\bibitem [{\citenamefont {Resta}(1998)}]{resta1998quantum}%
  \BibitemOpen
  \bibfield  {author} {\bibinfo {author} {\bibfnamefont {R.}~\bibnamefont
  {Resta}},\ }\href@noop {} {\bibfield  {journal} {\bibinfo  {journal}
  {Physical Review Letters}\ }\textbf {\bibinfo {volume} {80}},\ \bibinfo
  {pages} {1800} (\bibinfo {year} {1998})}\BibitemShut {NoStop}%
\bibitem [{\citenamefont {Jin}\ and\ \citenamefont {Song}(2019)}]{jin2019bulk}%
  \BibitemOpen
  \bibfield  {author} {\bibinfo {author} {\bibfnamefont {L.}~\bibnamefont
  {Jin}}\ and\ \bibinfo {author} {\bibfnamefont {Z.}~\bibnamefont {Song}},\
  }\href@noop {} {\bibfield  {journal} {\bibinfo  {journal} {Physical Review
  B}\ }\textbf {\bibinfo {volume} {99}},\ \bibinfo {pages} {081103} (\bibinfo
  {year} {2019})}\BibitemShut {NoStop}%
\bibitem [{\citenamefont {Levine}\ and\ \citenamefont
  {Steinhardt}(1984)}]{levine1984quasicrystals}%
  \BibitemOpen
  \bibfield  {author} {\bibinfo {author} {\bibfnamefont {D.}~\bibnamefont
  {Levine}}\ and\ \bibinfo {author} {\bibfnamefont {P.~J.}\ \bibnamefont
  {Steinhardt}},\ }\href@noop {} {\bibfield  {journal} {\bibinfo  {journal}
  {Physical review letters}\ }\textbf {\bibinfo {volume} {53}},\ \bibinfo
  {pages} {2477} (\bibinfo {year} {1984})}\BibitemShut {NoStop}%
\end{thebibliography}
\end{document}